\documentclass[reprint,amsmath,amssymb,aps]{revtex4-2}

\usepackage{graphicx}
\usepackage{dcolumn}
\usepackage{bm}
\usepackage{hyperref}
\usepackage{multirow}
\usepackage{braket}
\usepackage{xcolor}
\usepackage[draft]{changes}
\usepackage[normalem]{ulem}
\usepackage{cancel}

\begin{document}

\title{Evidence of a second-order phase transition in the six-dimensional Ising spin glass in a field}

\author{M. Aguilar-Janita}\affiliation{Complex Systems Group, Universidad Rey Juan Carlos, 28933 Móstoles, Madrid, Spain}
  
\author{V.~Martin-Mayor}\affiliation{Departamento de F\'\i{}sica
  Te\'orica, Universidad Complutense, 28040 Madrid,
  Spain}\affiliation{Instituto de Biocomputaci\'on y F\'{\i}sica de
  Sistemas Complejos (BIFI), 50018 Zaragoza, Spain}

\author{J.~Moreno-Gordo}\affiliation{Instituto de Biocomputaci\'on y
  F\'{\i}sica de Sistemas Complejos (BIFI), 50018 Zaragoza,
  Spain}\affiliation{Departamento de F\'\i{}sica Te\'orica,
  Universidad de Zaragoza, 50009 Zaragoza, Spain}\affiliation{Departamento de F\'{\i}sica,
  Universidad de Extremadura, 06006 Badajoz,
  Spain}\affiliation{Instituto de Computaci\'on Cient\'{\i}fica
  Avanzada (ICCAEx), Universidad de Extremadura, 06006 Badajoz,
  Spain}

\author{J.J.~Ruiz-Lorenzo}\affiliation{Departamento de F\'{\i}sica,
  Universidad de Extremadura, 06006 Badajoz,
  Spain}\affiliation{Instituto de Computaci\'on Cient\'{\i}fica
  Avanzada (ICCAEx), Universidad de Extremadura, 06006 Badajoz,
  Spain}\affiliation{Instituto de Biocomputaci\'on y F\'{\i}sica de
  Sistemas Complejos (BIFI), 50018 Zaragoza, Spain}

\date{\today}

\begin{abstract}
The very existence of a phase transition for spin glasses in an external magnetic field is controversial, even in high dimensions. We carry out massive simulations of the Ising spin-glass in a field, in six dimensions (which, according to classical ---but not generally accepted--- field-theoretical studies, is the upper critical dimension). We obtain results compatible with a second order phase transition and estimate its critical exponents for the simulated lattice sizes. The detailed analysis performed by other authors of the replica symmetric Hamiltonian, under the hypothesis of critical behavior, predicts that  the ratio of the renormalized coupling constants remain bounded as the correlation length grows. Our numerical results are in agreement with this expectation. 
\end{abstract}

\maketitle
\section{Introduction}
The existence of a spin-glass (SG) transition in the presence of an
external magnetic field, at the so-called de Almeida-Thouless (dAT)
line~\cite{dealmeida:78}, is one of the most challenging problems in the
realm of disordered systems~\cite{PVM_book,young_book,VEHertz,VEBovier}.  The existence of
the dAT line is firmly established only in the limit of infinite space
dimensions, $D\to\infty$~\cite{PVM_book}.

In order to clarify this problem, the community has tried to implement
the Wilson renormalization group (RG)
program~\cite{wilson:74,Amit-Martin}. The starting point is
the computation of the so-called upper critical-dimension $D_\text{u}$
(the smallest $D$ at which critical exponents take their Gaussian
values).  Unfortunately, this approach, extremely successful in an enormous
variety of problems including SG at zero
field~\cite{harris:76,janus:13}, has not yet succeed.
Replicated field-theory finds $D_\text{u}=6$, but fails to
find, in the one-loop approximation, a fixed point stable at $D=6$
~\cite{bray:80,pimentel:02}.  In fact, we identify no
fewer than eight conflicting scenarios:

(i) The droplets model predicts the absence of a dAT line for any
$D<\infty$ (i.e., the so-called \emph{lower} critical dimension is
$D_{\text{l}}\!=\!\infty$)~\cite{mcmillan:84,fisher:86,bray:87,fisher:88}.

(ii) A modified version of the droplets model, finds
$D_{\text{u}}=D_{\text{l}}=6$. In other words, a spin-glass phase
transition would be possible in a field only if $D>6$~\cite{yeo:15}. See also
Refs.~\cite{bray:11} and \cite{parisi:12}.

(iii) An analysis based in the study of high-temperature series finds a second phase transition only for $D\ge 6$~\cite{singh:17b}.

(iv) A very recent field-theoretical analysis claims
$D_{\text{u}}\!=\!8$~\cite{Angelini:22}. 

(v) Interestingly enough, a two-loop computation
does find a nontrivial stable fixed point at
$D=6$~\cite{charbonneau:17,charbonneau:19}, the Gaussian one being unstable. 
This nontrivial fixed-point would lie in the
nonperturbative region (which makes it unclear whether or not the
fixed point would survive a three-loops computation). 

(vi) The scenario described in Ref.~\cite{holler:20} predicts a quasi-first order transition in a field.

(vii) Large scale numerical simulations suggest the presence of a dAT line for $D\!=\!4$~\cite{janus:12}, but the results are not conclusive for $D\!=\!3$~\cite{janus:14b,janus:14c}.

(viii) The study of $D\!=\!1$ models with long-range interactions that mimic short-range models at $D\!>\!1$ provides somewhat contradicting results. Some studies argue that there must not be a dAT line below $D\!=\!6$~\cite{katzgraber:05b,katzgraber:09,vedula:23}, while others claim in favor of a dAT line for much lower $D$~\cite{leuzzi:09,dilucca:20}.

Here, we add clarity to the debate by showing numerical evidences that a dAT line is
present in $D\!=\!6$ through massive numerical simulations: our largest
lattices contain $8^6$ spins, more than twice the $48^3$ spins in the largest system  ever equilibrated in $D\!=\!3$~\cite{fernandez:09b}.
We find that the scaling behavior of the different susceptibilities is qualitatively different from their zero field counterpart. We have also been able to estimate the critical exponents. However, the lack of analytical predictions for the logarithmic-corrections of the different observables and the size of our statistical errors prevent us from making any strong claim regarding the value of $D_{\text{u}}$. Finally, we obtain numerical evidences that  the two-parameter replica symmetric (RS) effective Hamiltonian \cite{pimentel:02} is able to describe the scaling of the ratio of the renormalized coupling constants in the critical region, which enable us to discuss the problem of a second order phase transition versus a quasi-first order one.

The structure of the paper is as follows. First, we describe our model (Sect. \ref{sec: Model}), the field-theoretical framework (Sect.~\ref{subsect:Field-Theoretical}) and the finite size scaling techniques (Sect.~\ref{subsect:FSS}) used to study it. Then, we present our numerical results in Sect.~\ref{sec:results}. In particular, in this section we briefly introduce the details of our simulations (Sect.~\ref{subsec:simulations}), we study the replicon and anomalous susceptibilities (Sect. \ref{subsec:suscep}), the probability density function of the overlap (Sect. \ref{sec:overlap_pdf}), the critical exponents (Sect.~\ref{subse:critical}), and the parameter $\lambda$ (Sect.~\ref{subsec:lambda}). Finally, we discuss our results and draw some conclusions in Sect. \ref{sec:conclusion}. 

The paper is supplemented by eight appendixes, organized as follows. In Appendix~\ref{appA} we present the effective Hamiltonian and the structure of the propagators. In Appendix~\ref{app:Hemholtz-Gibbs} we revisit theoretical results pertaining to the Helmholtz and Gibbs free energies. Appendix~\ref{app:lambda-def} is dedicated to defining the parameter $\lambda_r$ and the study of its estimators. Details of our simulations and algorithms are presented in Appendix~\ref{app:details} and Appendix~\ref{Multi-spin-Apendice}. Appendix~\ref{app:quotients} address the computation of the critical exponents. In Appendix~\ref{H0_appendix} we present numerical results for the $h=0$ phase transition of the six dimensional Ising spin glass. We conclude by studying the $\Lambda$ cumulants on Appendix~\ref{app:Lambda}.

\section{The Model}\label{sec: Model}
 We consider the Edwards-Anderson Hamiltonian for Ising spins (i.e., $s_{\boldsymbol{x}}=\pm1$) on a six dimensional cubic lattice of size $V=L^6$, with periodic boundary conditions and nearest-neighbors interactions: 
\begin{equation}\label{hamiltonian}
    \mathcal{H} = - \sum_{\langle \boldsymbol{x}, \boldsymbol{y}\rangle} J_{\boldsymbol{x} \boldsymbol{y}} s_{\boldsymbol{x}} s_{\boldsymbol{y}}
    -h\sum_{\boldsymbol{x}} s_{\boldsymbol{x}}\;,
\end{equation}
where the couplings are independent, identically distributed random
variables ($J_{\boldsymbol{x} \boldsymbol{y}}=\pm 1$ with equal
probability). Hereafter, the over line $\overline{(\cdots)}$ means
average over the couplings, and $\langle(\cdots)\rangle$ is the
thermal average carried-out for fixed couplings $\{J_{\boldsymbol{x}
  \boldsymbol{y}}\}$.  A choice of couplings is named a \emph{sample}.
 
\subsection{Field-theoretical framework}\label{subsect:Field-Theoretical}
 The analysis of the RS Hamiltonian in the field theory (see Appendix~\ref{appA}) finds
 three masses (replicon, anomalous and longitudinal) and their
 associated propagators (correlation functions)~\cite{pimentel:02,parisi:13}.
 In the spin glass phase, the most singular mode
 is the replicon. The anomalous and longitudinal modes become
 identical when one takes the limit for the number of replicas $n$
 going to zero. Hence the two fundamental propagators of the theory, $
 G_R(\boldsymbol{x}-\boldsymbol{y})$ and $
 G_{\text{A}}(\boldsymbol{x}-\boldsymbol{y})$, are defined as (see Appendix~\ref{appA} for more details)
\begin{equation}\label{eq:GR}
G_R(\boldsymbol{x}-\boldsymbol{y})= \overline{ \langle s_{\boldsymbol{x}} s_{\boldsymbol{y}} \rangle^2} -2
 \overline{ \langle s_{\boldsymbol{x}} s_{\boldsymbol{y}} \rangle \langle s_{\boldsymbol{x}}\rangle \langle s_{\boldsymbol{y}} \rangle}
 +  \overline{ \langle s_{\boldsymbol{x}} \rangle ^2 \langle s_{\boldsymbol{y}} \rangle^2}\;
\end{equation}
and
\begin{equation}\label{eq:GA}
 G_{\text{A}}(\boldsymbol{x}-\boldsymbol{y})= \overline{ \langle s_{\boldsymbol{x}} s_{\boldsymbol{y}} \rangle^2} -4
 \overline{ \langle s_{\boldsymbol{x}} s_{\boldsymbol{y}} \rangle \langle s_{\boldsymbol{x}}\rangle \langle s_{\boldsymbol{y}} \rangle}
 +3  \overline{ \langle s_{\boldsymbol{x}} \rangle ^2 \langle s_{\boldsymbol{y}} \rangle^2}\;.
\end{equation}

Associated with each two-point correlation functions one can define a susceptibility $\chi$  as 
\begin{equation}\label{suscep}
    \chi_{\alpha}=\widehat{G}_{\alpha}(\boldsymbol{0})\; \;\;\; \alpha\in \{R,A\}\;,
\end{equation}
where $\widehat{G}(\boldsymbol{k})$ is the discrete Fourier transform of $G(\boldsymbol{x})$ (see Eq. (\ref{eq:Fourier})). In the $h=0$ case, it is straightforward to show that $\chi_{\text{R}}=\chi_{\text{A}}=\chi_{\text{L}}$. When $h\ne 0$ instead, we shall find below that $\chi_{\text{R}}$ becomes dominant in the spin glass phase.

In order to study the RS effective Hamiltonian, one introduces
$\omega_1$ and $\omega_2$, which are the following three-point connected
correlation functions at zero external momentum 
\begin{equation}\label{eq:def-omega1}
    \omega_1= \frac{1}{V}\sum_{\boldsymbol{x} \boldsymbol{y} \boldsymbol{z}} \overline{\langle s_{\boldsymbol{x}} s_{\boldsymbol{y}}\rangle_c \langle s_{\boldsymbol{y}} s_{\boldsymbol{z}}\rangle_c \langle s_{\boldsymbol{z}} s_{\boldsymbol{x}}\rangle_c}\;,
\end{equation}
\
\begin{equation}\label{eq:def-omega2}
    \omega_2= \frac{1}{2V}\sum_{\boldsymbol{x}\boldsymbol{y}\boldsymbol{z}}\overline{\langle s_{\boldsymbol{x}} s_{\boldsymbol{y}} 
    s_{\boldsymbol{z}}\rangle^2_c}\;,
\end{equation}
where the $c$ subindices stand for \emph{connected} correlation
functions. We refer the reader to Appendixes ~\ref{app:Hemholtz-Gibbs} and ~\ref{app:lambda-def} for the technical details about the computation of $\omega_1$ and $\omega_2$.

An interesting observable is the ratio of the two renormalization vertices of the theory, denoted $w_{1,r}$ and $w_{2,r}$
\begin{equation}
    \lambda_r = \frac{w_{2,r}}{w_{1,r}}\,.
\end{equation}
Interestingly enough, this ratio can be easily obtained as well (see Appendix ~\ref{app:lambda-def} for additional details) in terms of $\omega_1$ and $\omega_2$ as

\begin{equation}
 \lambda_r = \frac{\omega_2}{\omega_1}\,. \label{eq:lambda-calculable}
\end{equation}
This equation allows us to compute the ratio of the two renormalized couplings ($\lambda_r$) in a numerical simulation on the lattice by computing the quotient of two connected correlations functions at zero external momentum.

Finally, let us remark that there are two 
different ways of taking the limits for $\lambda_r(L,T)$ at $T_c$
\begin{equation}
    \lambda_r^* = \lim_{L\to\infty}\lim_{T\to T_c} \lambda_r(L,T)\,,\ \lambda_r(T_c^{+}) = \lim_{T\to T_c^+}\lim_{L\to\infty} \lambda_r(L,T)\,.
\end{equation}
In principle, $\lambda_r^* \ne  \lambda_r(T_c^{+})$. We are interested in  $\lambda_r(T_c^{+})$ \footnote{In order to compute $\lambda_r^*$ at the upper critical dimension one could try to adapt to the RS Hamiltonian the strategy followed by Brezin and Zinn-Justin for the $\phi^4$ theory~\cite{zinn-justin:05}. In particular, the theory would be studied from the beginning on a finite box, exactly at its infinite-volume critical temperature. We are not aware of any such computation, not even in the simplest case of the one-component $\phi^3$ theory.}.

Notice that six real replicas (six independent copies of the system evolving under the same couplings) are
needed to compute $\omega_1$ and $\omega_2$. However, one can compute $\omega_1$ and $\omega_2$ in terms of
three and four real-replica estimators (see Appendix~\ref{app:lambda-def}).  Within the framework of the RS
theory~\cite{parisi:13}, the values of the three and four-replica estimators differ in
general from the true values of $\omega_1$ and $\omega_2$  but coincide with them at the critical
temperature. This gives us the opportunity to check the validity of
the RS theory by computing the six, four and three-replica
estimators~\footnote{In Ref.~\cite{fernandez:22} the
four-dimensional Ising spin glass in a field was analyzed numerically
but $\omega_1$ and $\omega_2$ were computed only from three- and four-
replica estimators.}.

Another check is the value of $\lambda_r$ itself, since the replica
symmetric field theory predicts a value of $0\le\lambda_r\le1$ for a second-order phase transition
while a
value of $\lambda_r > 1$ would imply the presence of a
quasi-first-order phase transition~\cite{holler:20}. It is worth
noting that $\lambda_r$ controls as well the mean-field (MF) values of equilibrium and
off-equilibrium dynamical exponents~\cite{parisi:13}.

\subsection{Finite Size Scaling} \label{subsect:FSS}
We want to investigate whether or not the systems undergoes a second-order
phase transition in the presence of a magnetic field and, if the
answer is positive, to characterize the resulting  universality class. Indeed, a standard
way of identifying a phase transition is computing  some correlation length $\xi$, that is used to identify scale invariance. An appropriate definition of the
second-moment correlation length in a finite lattice is~\cite{Amit-Martin}
\begin{equation}\label{xi2}
    \xi_2 = \frac{1}{2\sin(\pi/ L)}\left(\frac{\widehat{G}_R(0)}{\widehat{G}_R(\boldsymbol{k_1})}-1\right)^{1/2}\, ;
\end{equation}
where $\boldsymbol{k_1} = (2\pi/L,0,0,0,0,0)$ (or permutations). The scale invariance of $\xi_2/L$ at the critical point results in
\begin{equation}\label{scalingxi}
    \frac{\xi_2}{L} = f_{\xi}(L^{\frac{1}{\nu}}t)+ L^{-\omega} g_{\xi}(L^{\frac{1}{\nu}}t) + \dots \, ,
\end{equation}
where $\omega$ is the correction-to-scaling exponent and
$t=(T-T_\mathrm{c})/T_\mathrm{c}$ is the reduced temperature. From this behavior one
expects that a plot of $\xi_2(T)/L$ for several system sizes will show a
common intersection point at $T=T_\mathrm{c}$, provided that the sizes are large
enough to make corrections to scaling negligible. However, previous works in lower
dimensions~\cite{janus:12,leuzzi:09} did not find this intersection.  This anomalous behavior was
attributed to an abnormal behavior of the propagator at wave vector
$\boldsymbol k=\boldsymbol 0$~\cite{leuzzi:09}, that induces strong corrections to
the leading scaling-behavior in Eq. (\ref{scalingxi}). This phenomenon
is illustrated through the spin-glass order parameter distribution in Sec.~\ref{sec:overlap_pdf}.

Given the aforementioned anomaly, we consider a second scale-invariant
quantity, previously introduced in Ref.~\cite{janus:12} under the name
$R_{12}$, which  is computed as a dimensionless ratio of propagators with higher
momenta
\begin{equation}\label{eq:R12}
    R_{12}=\frac{\widehat{G}_R\left(\boldsymbol{k}_{1}\right)}{\widehat{G}_R\left(\boldsymbol{k}_{2}\right)}\;.
\end{equation}
Here, $\boldsymbol{k}_{1}$ and $\boldsymbol{k}_{2}$ are the smallest nonzero momenta compatible with periodic boundary conditions, namely
$\boldsymbol{k}_{1} = (2\pi/L,0,0,0,0,0,0)$ and $\boldsymbol{k}_{2}= (2\pi/L,\pm 2\pi/L,0,0,0,0)$ (and permutations). Notice that $R_{12}$ scales in the same way as $\xi_2/L$; see Eq.~\eqref{scalingxi}.

\section{Numerical results}\label{sec:results}
In this section we present the numerical results obtained from our simulations. We begin by briefly describing our simulations. Subsequently, we delve into the study of the susceptibilities, the probability density function of the overlap, the critical exponents and the parameter $\lambda_r$. 
\subsection{Description of simulations}\label{subsec:simulations}
We have studied the model in Eq.~\eqref{hamiltonian} through Monte Carlo simulations on lattices $L=5,6,7$ and $8$, with a magnetic field set to $h=0.075$.  Thermalization is ensured by using  the parallel tempering
algorithm~\cite{hukushima:96,marinari:98b}, complemented with a demanding equilibration test based on Ref.~\cite{billoire:18}. In order to obtain high statistics, we have simulated 25600 samples for $L=5,6,7$ and 5120 for the largest lattice size $L=8$ by using multispin coding. Six statistically independent system copies of each sample, named real replicas, are simulated
in order to compute without statistical bias both $\omega_1$ and $\omega_2$; recall Eqs.~\eqref{eq:def-omega1} and~\eqref{eq:def-omega2}. Further details about
our simulations are provided in Appendixes ~\ref{app:details} and ~\ref{Multi-spin-Apendice}.

\subsection{Replicon and Anomalous susceptibilities}\label{subsec:suscep}

\begin{figure}
    \centering
    \includegraphics[scale=0.68]{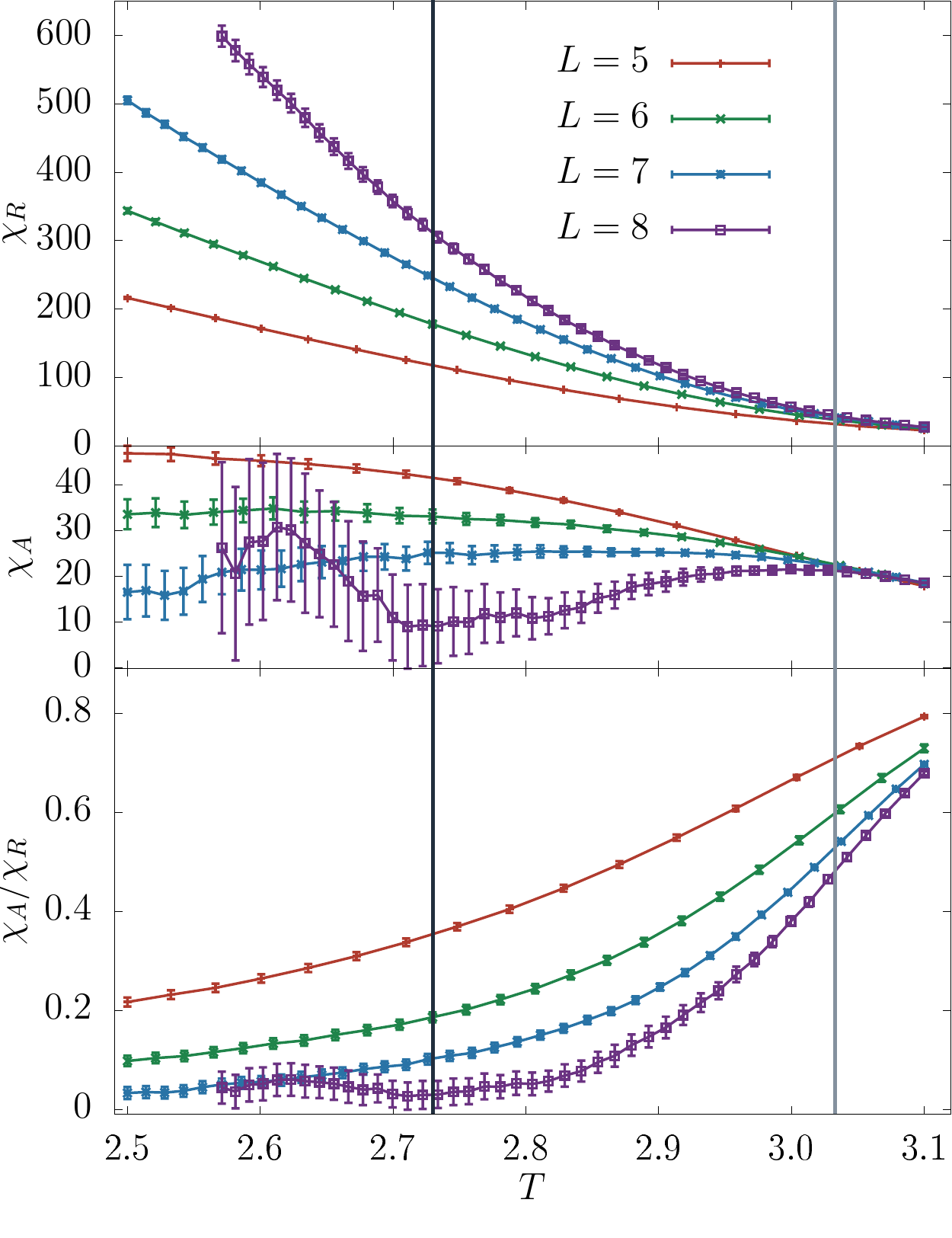}
    \caption{Susceptibilities for the replicon $\chi_{\text{R}}$ (on the {\bf top}), Eqs.~\eqref{eq:GR} and~\eqref{suscep}, for the anomalous mode $\chi_{\text{A}}$ (on the {\bf middle}), Eqs.~\eqref{eq:GA} and~\eqref{suscep}, and for the ratio of the anomalous mode and the replicon one, $\chi_{\text{A}}/\chi_{\text{R}}$, (on the {\bf bottom}), vs temperature $T$, as computed for our different system sizes in a magnetic field $h=0.075$. For $h=0$ one trivially shows that $\chi_{\text{R}}=\chi_{\text{A}}$. Instead, for $h=0.075$, we find that $\chi_{\text{R}}$ rapidly grows with $L$ at low temperatures, while $\chi_{\text{A}}/\chi_{\text{R}}$ goes to zero. The different size dependence ensures that we are working far enough from the $h=0$ point in the dAT line. The two vertical lines are our estimates for the critical temperatures for $h=0.075$ (left vertical line) and for $h=0$ (right line). In particular, note from the middle panel that the values of $\chi_{\text{A}}$, as computed in $L=7$ and 8 lattices, are compatible: i.e., differences smaller than two standard deviations at, and below, our estimated critical temperature for $h=0.075$.}
    \label{fig:susRysusA}
\end{figure}

One may question if our  magnetic field $h=0.075$ is large enough to ensure that we are working far-enough from the $h=0$ endpoint of the dAT line. We answer that question by computing the replicon and anomalous susceptibilities: notice that for $h=0$, $\chi_{\text{A}}=\chi_{\text{R}}$. In Fig. \ref{fig:susRysusA} we represent $\chi_{\text{R}}$ (top panel), $\chi_{\text{A}}$ (middle panel), and the ratio $\chi_A/\chi_R$ (bottom panel). Indeed, at the critical point even our smallest system $L=5$ has $\chi_{\text{R}}(L=5,T_\mathrm{c})\approx$ 3 $\chi_{\text{A}} (L=5,T_\mathrm{c})$, and this ratio gets larger as $L$ grows (see bottom panel in Fig. \ref{fig:susRysusA}) \footnote{One can compare this factor with the values of $\chi_{R}/\chi_{A}$ at $T_c(h=0)$, which are $\chi_{R}/\chi_{A}\simeq 1.35, 1.65, 1.85, 1.96$ for $L=5, 6, 7 $ and $8$ respectively}, meaning that correlations extend to a much longer distance for the replicon mode than for the anomalous one, in agreement with both the MF picture and previous computations in $D=3$~\cite{janus:14b} and $D=4$~\cite{fernandez:22}.

\subsection{Probability density function of the overlap}\label{sec:overlap_pdf}

\begin{figure}[h]
    \centering
    \includegraphics[scale=0.59]{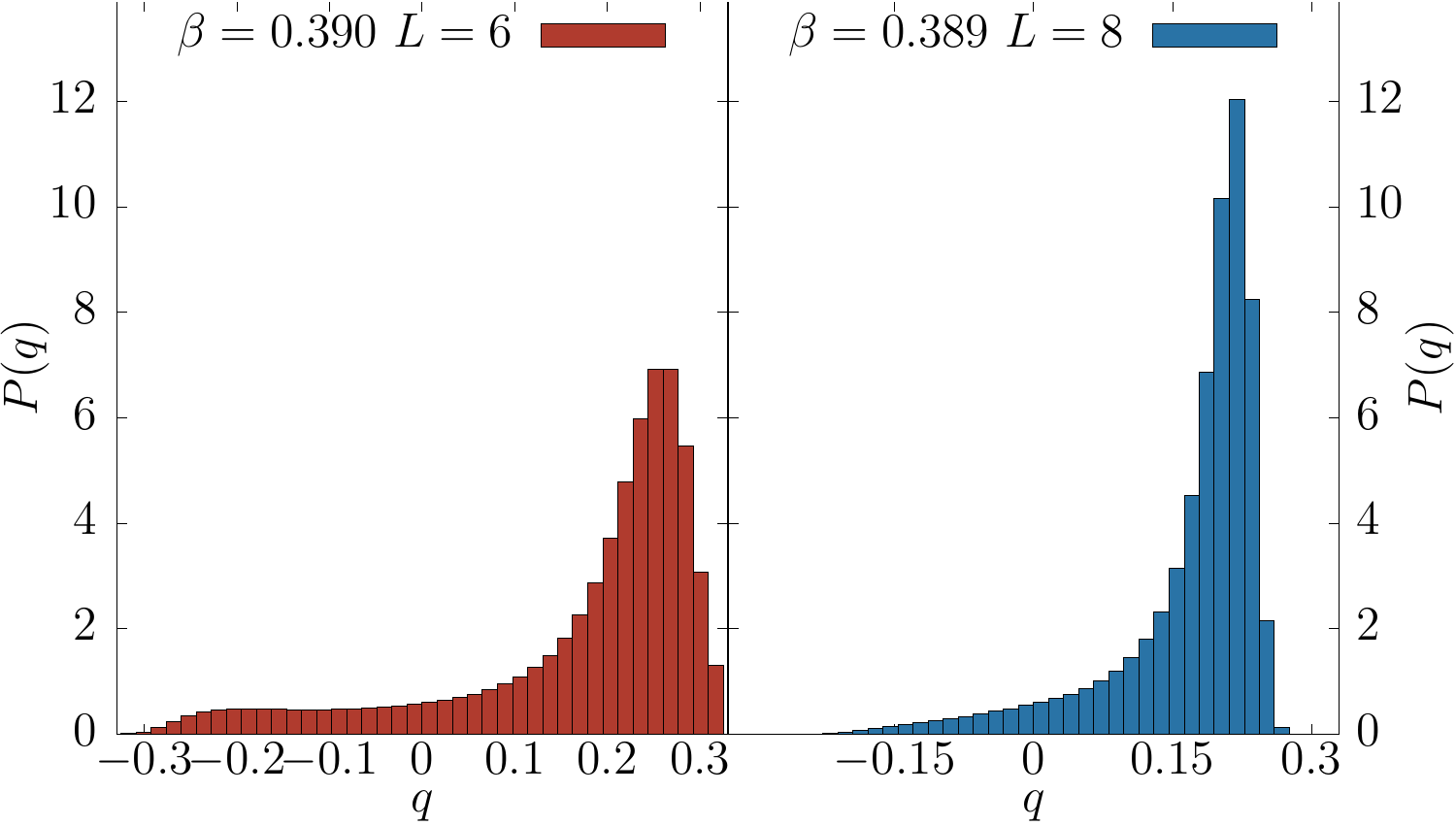}
    \caption{Probability density function of the overlap $P(q)$ for two different sizes: $L=6$ (left) and $L=8$ (right). We have computed $P(q)$ for the lowest available temperature for $L=8$ ($\beta \approx 0.389$) and the closest available temperature for $L=6$ ($\beta \approx 0.390$).}
    \label{fig:histograma_Pq}
\end{figure}

We have computed the probability density function of the overlap, $q$, for $L=6$ and $L=8$. We observe (see Fig.~\ref{fig:histograma_Pq}) for both lattice sizes that the probability density function is nonzero for negative overlaps. Note that the tail at $q<0$ is suppressed in the thermodynamic limit. Indeed, the trend towards a  suppression of the tail as $L$ grows is very clear from our data in Fig.~\ref{fig:histograma_Pq}.

As mentioned in the previous section, the nonnegligible tail of negative overlaps is probably responsible of the undesirable behavior of the propagator at wave vector $\boldsymbol{k}=\boldsymbol{0}$, which makes it difficult to find intersections of the curves corresponding to different (small) lattice sizes of
$\xi_2/L$ as a function of temperature. 

\subsection{Critical exponents}\label{subse:critical}
\begin{figure}
    \centering
    \includegraphics[scale=0.68]{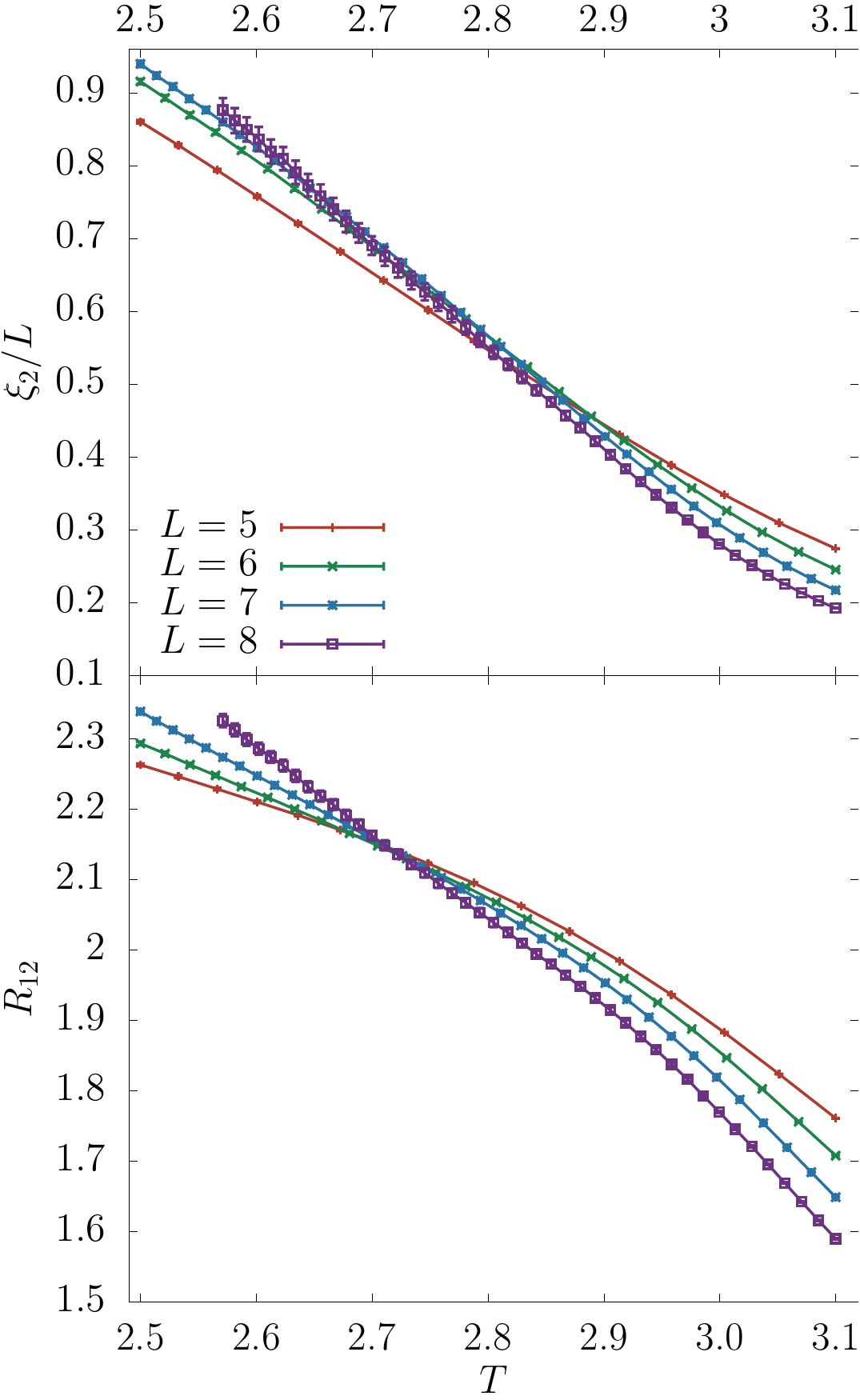}
    \caption{Second-moment correlation length $\xi_2$, Eq.~\eqref{xi2}, measured in units of the lattice size $L$  (on the {\bf top}),
and dimensionless ratio $R_{12}$, Eq.~\eqref{eq:R12} (on the {\bf bottom}), as a function of temperature $T$. Both quantities are shown
    as computed for the replicon propagator, Eq.~\eqref{eq:GR}, for all our lattice sizes at $h=0.075$. At the critical point, the curves for different sizes of the system intersect at the same $T$, meaning
    that both $\xi/L$ and $R_{12}$ are a scale invariant at $T_c$. The presence of a crossing point for $\xi_2/L$  indicates that the anomalous behavior of the wave vector $\boldsymbol{k}=\boldsymbol{0}$ is less severe in space dimension $D=6$ than previously found at $D=4$~\cite{janus:12}. The difference between the crossing points found for both quantities
    should vanish as $L\to\infty$, since it is due to scaling corrections.}
    \label{fig:corte_xi}
\end{figure}

We start by determining the value of the critical temperature at which the phase transition takes place. Eq.~\eqref{scalingxi} tells us that the curves of dimensionless magnitudes such as $\xi_2/L$ and $R_{12}$, when computed for different system sizes, will intersect at $T_c(h)$. These intersections are shown in Fig. \ref{fig:corte_xi} and in Table \ref{tabla_beta_c}. Corrections to scaling cause the intersection
points to vary, depending on the considered pair of lattice ($L_1,L_2$) and the
quantity under inspection, $\xi_2/L$ or $R_{12}$. However, for our largest systems
($L_1\!=\!7$, $L_2\!=\!8$) we find compatible crossings for $\xi_2/L$ and $R_{12}$. Also the ($L_1\!=\!6$, $L_2\!=\!7$) crossing for $R_{12}$ turns out to be compatible with the results from ($L_1\!=\!7$, $L_2\!=\!8$).  Although a more accurate estimation
 of $T_\mathrm{c}$ is obtained below, the reader can already appreciate that $T_\mathrm{c}(h=0.075)$ is significantly smaller than its $h=0$ counterpart $T_\mathrm{c}(h=0)=3.033(1)$~\cite{wang:93,klein:91}.  See also Appendix \ref{H0_appendix}.
 
\begin{table}[b]
\begin{tabular}{cccc} \hline \hline
$L_1$ & $L_2$ & $T_c^{\xi}$ & $T_c^{R}$ \\ \hline
5 & 6 & 2.892(6) &2.680(14)             \\
6 & 7 & 2.809(16) &2.739(15)          \\ 
7 & 8 & 2.69(4) & 2.74(2)      \\ \hline \hline
\end{tabular}
\caption{Temperatures for the crossing points of $\xi/L$ and  $R_{12}$ for consecutive sizes.}
\label{tabla_beta_c}
\end{table}

Our next step is the characterization of the universality class by computing  
critical exponents $\nu$
(associated with the correlation length) and $\eta$ (associated with the replicon susceptibility). We extract effective, size-dependent exponents by using the quotient method~\cite{nightingale:76,ballesteros:96b,ballesteros:97} (see Table \ref{table:exponents}). To avoid the  somewhat problematic $ \boldsymbol{k}=\boldsymbol{0}$ wave vector we compute $\nu$ from the scaling of $\partial R_{12}/\partial T$ (see Appendix~\ref{app:quotients}) and $\eta$ from the susceptibility $\mathcal{F}$:
\begin{equation}
    \mathcal{F} = \widehat{G}_R(\boldsymbol{k_1}) \,. 
\end{equation}

\begin{table}[b]
\begin{tabular}{cccc}
\hline \hline 
$L_1$ & $L_2$ & $\eta(T_c^{R})$ & $\nu(T_c^{R})$ \\ \hline 
5   & 6     & 0.40(1)    & 0.88(7) \\ 
6   & 7     & 0.28(1)    & 0.76(5) \\ 
7   & 8     & 0.21(1)    & 0.54(6)  \\ \hline \hline
\end{tabular}
\caption{Effective exponents $\eta$ and $\nu$ as obtained from the quotient method for lattices $(L_1,L_2)$. Values are shown
at the temperatures obtained from the crossings in $R_{12}$.}
\label{table:exponents}
\end{table}

The effective exponents in Table~\ref{table:exponents} need to be extrapolated to
$L_1\to\infty$. We have checked that these extrapolations are compatible
with  MF values,  $\nu=1/2$ and
 $\eta=0$.  We have started with independent fits to the laws $\nu(L_1)=1/2+O(L_1^{-\omega_\nu})$ and
 $\eta(L_1)=O(L_1^{-\omega_\eta})$ obtaining good fits, where $\omega_i$ ($i=\nu, \eta$) is the leading correction-to-scaling exponent\footnote{Considering only leading corrections to scaling, the quotient of $\partial R_{12}/\partial T$ at the crossing point for ($L_1,L_2)$ is $Q=(L_2/L_1)^{1/\nu}[1+A_1 L_2^{-\omega}]/[1+A_1 L_1^{-\omega}]$, where $A_1$ is an amplitude and $\omega$ is the leading corrections-to-scaling exponent. If $L_2=L_1+1$, $Q=(L_2/L_1)^{1/\nu}[1+A_1 (L_2^{-\omega}-L_1^{-\omega})+\ldots]$. Thus, the exponents we are computing from the fits are $\omega_{\nu,\eta}=\omega+1$.}. For $\nu$ 
 we get $\chi^2/\mathrm{dof}=2.9/1$ (dof is the number of degrees of
 freedom) with a $p$-value=$9\%$ and $\omega_\nu=3.7(1.3)$. For 
 $\eta$ we obtain $\chi^2/\mathrm{dof}=0.03/1$ with a $p$-value=$87\%$ and
 $\omega_\eta=1.93(15)$. We can improve by trying a
 joint fit for $\nu$ and $\eta$ that assumes a common value of $\omega_{\nu\eta}$ (in agreement with the RG expectation). The joint fit 
  obtains $\omega_{\nu \eta}=1.96(15)$ with
 $\chi^2/\mathrm{dof}=5.67/3$ with a $p$-value=$13\%$. Our extrapolations to large $L_1$ not constrained to yield MF exponents resulted in exceedingly large errors for both $\nu$ and $\eta$. Thus, although our estimated exponents are compatible with an upper critical--dimension in a field $D^h_{\text{u}}=6$, we cannot exclude nearby values for $D^h_{\text{u}}$. In particular, the largeness of the $\omega$ exponent seems in contradiction with the logarithmic corrections (i.e., $\omega = 0$) expected at the upper critical dimension.
 
At this point, we are ready for a joint extrapolation to infinite system sizes of the critical temperature. With the data appearing in Table I, we perform a join fit with the two data sets to
\begin{equation}
    \beta^L_c = \beta^\infty_c + A \, \dfrac{s^{-\omega} - 1}{1-s^{1/\nu}} L^{-\omega - 1/\nu} \, ,
\end{equation}
where $\beta = 1/T$, $s=(L+1)/L$,  $\omega=5(2)$, and $\nu$ is fixed to $\nu=1/2$. The results of the fit are $T_c=1/\beta^\infty_c=2.755(13)$ with $\chi^2/\mathrm{dof}=5.02/2$ with a $p$-value=$8.0\%$.

We have also checked that the impact of the factor $f(s)=(s^{-\omega} - 1)/(1-s^{1/\nu})$ is negligible. Actually, if we perform the same fit setting $f(s)=1$ we obtain: $T_c=1/\beta^\infty_c=2.755(15)$ with $\chi^2/\mathrm{dof}=5.48/2$ with a $p$-value=$6.5\%$.

 \subsection{Study of the $\lambda_r$ parameter}\label{subsec:lambda}
Finally, let us consider the $\lambda_r$ parameter from Eq.~\eqref{eq:lambda-calculable}, that we have computed using the three-, four-, and six-replica estimators of $\omega_1$ and $\omega_2$. These multiple calculations enable us to check two predictions from the RS theory. First, it predicts that the three and four replicas estimators give the true value of $\lambda_r$ (which is the one of the six-replica estimator) near the critical point. Second, this theory predicts a value of $\lambda_r\in [0,1]$. The two following subsections address these points. 

\subsubsection{Comparison of $\lambda_r$ computed with three, four and six replicas}
\begin{figure}[h!]
    \centering
    \includegraphics[scale=0.68]{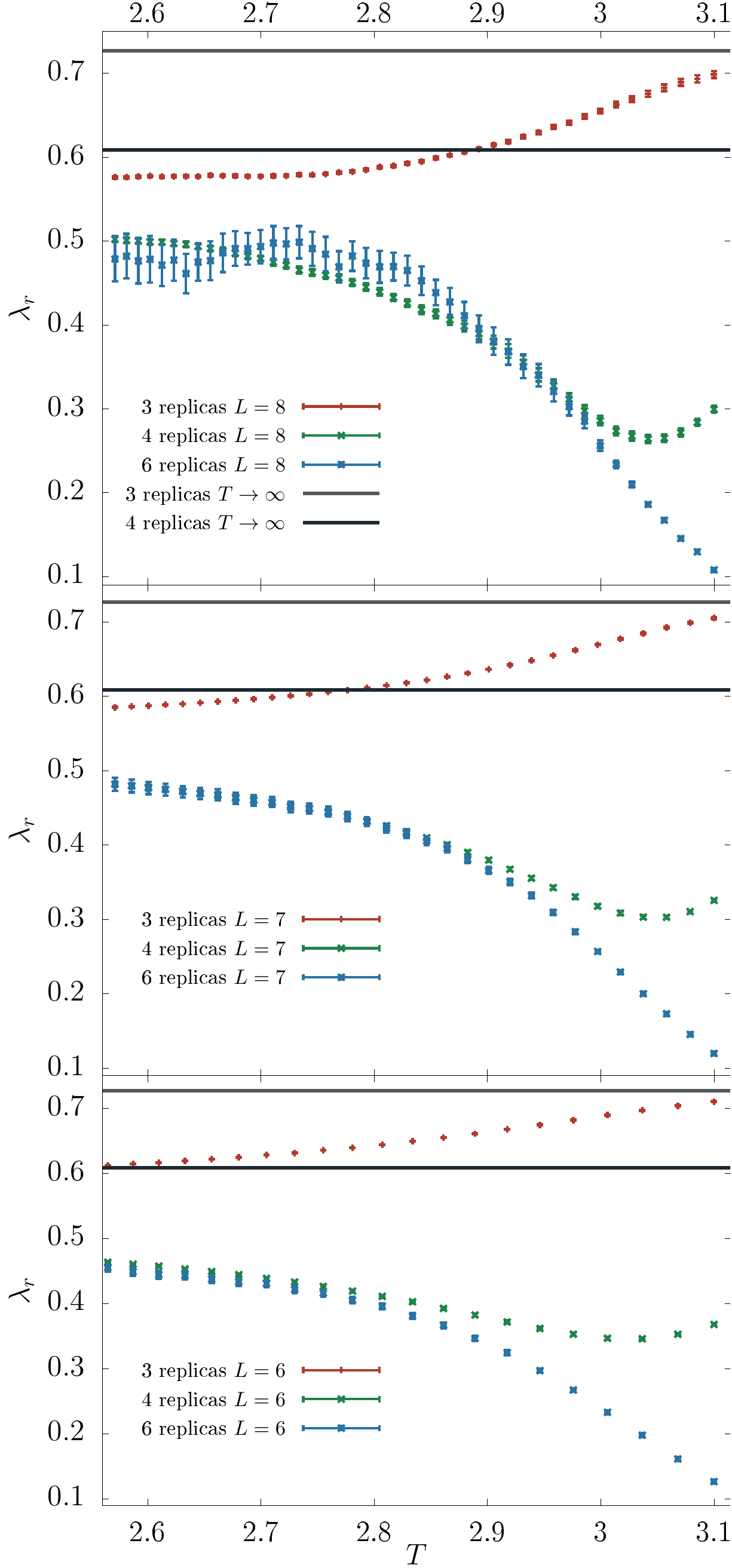}
    \caption{Plot of the three-, four- and six-replica estimator of $\lambda_r$, Eqs. \eqref{eq:omega_3replicas}, \eqref{eq:omega_4replicas} and \eqref{eq:omegas} respectively, as a function of the temperature for $L=6$ (bottom panel), for $L=7$ (middle panel), and for $L=8$ (top panel). We also plot the infinite temperature limits for three and four replicas.}
    \label{fig:lambda346_L8}
\end{figure}

We represent in Fig.~\ref{fig:lambda346_L8} the three-, four-, and six-replica estimator of $\lambda_r$ for different lattice sizes $L=6,7,$ and $8$ (bottom, middle and top panels respectively). We also plot the infinite temperature values for three and four replicas. The three-replica estimator seems to converge to the infinite value faster than the four-replica estimator, which exhibit a crossover between the six-replica behavior at low temperatures and the infinite temperature one at higher temperatures. Note that, at the estimated critical temperature $T_c = 2.755(13)$, the four- and six-replica estimators are compatible within one standard deviation. Thus, we will use the four-replica estimator which exhibits a smaller statistical error.

\begin{figure}[h]
    \centering
    \includegraphics[scale=0.68]{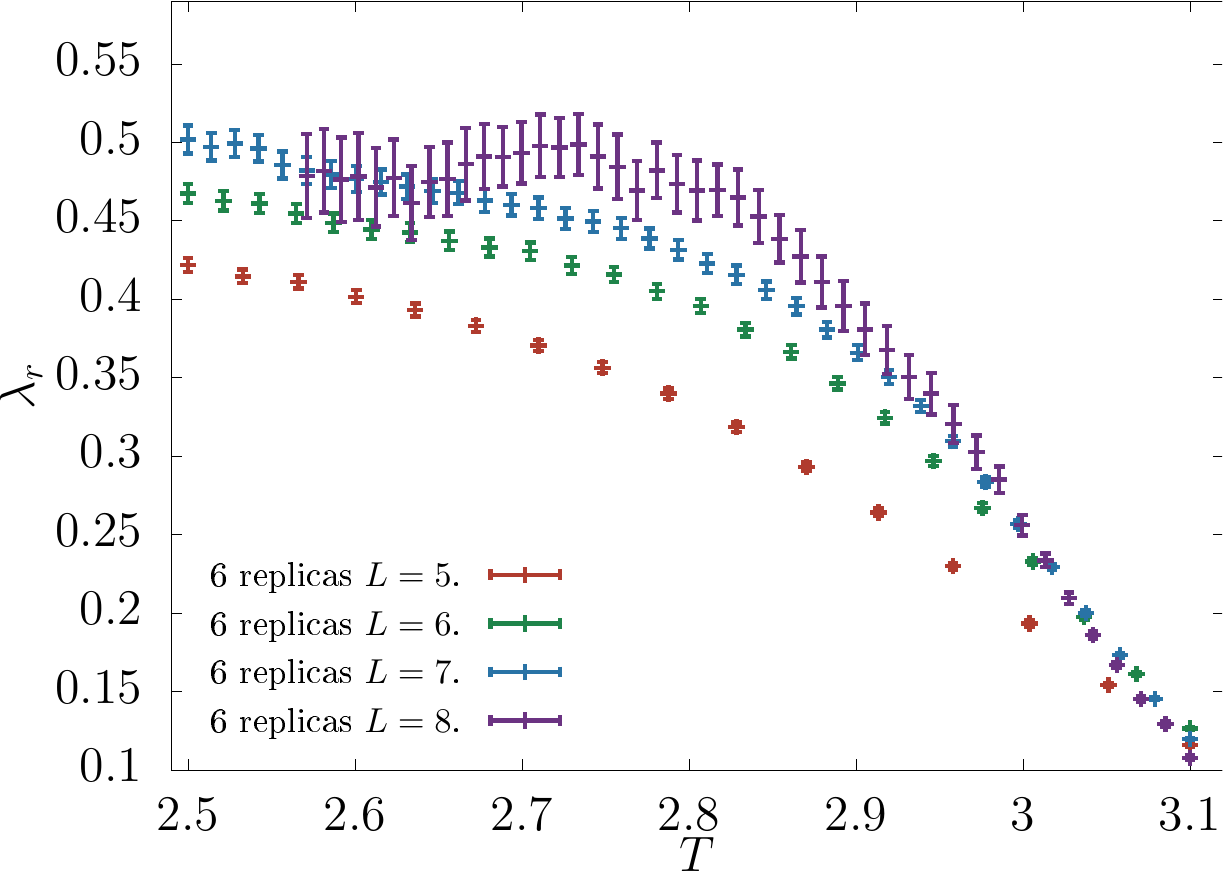}
    \caption{Plot of the six-replica estimator of $\lambda_r$, Eq. \eqref{eq:omegas} and \eqref{eq:lambda_omegas}, as a function of the temperature for all the lattice sizes $L$.}
    \label{fig:lambda6}
\end{figure}

In Fig.~\ref{fig:lambda6} we show the values of six-replica estimator of $\lambda_r$ as a function of the temperature for different lattice sizes.

\subsubsection{Value of \(\lambda_r(T_c^+)\)} 
\begin{figure}[h]
    \centering
    \includegraphics[scale=0.7]{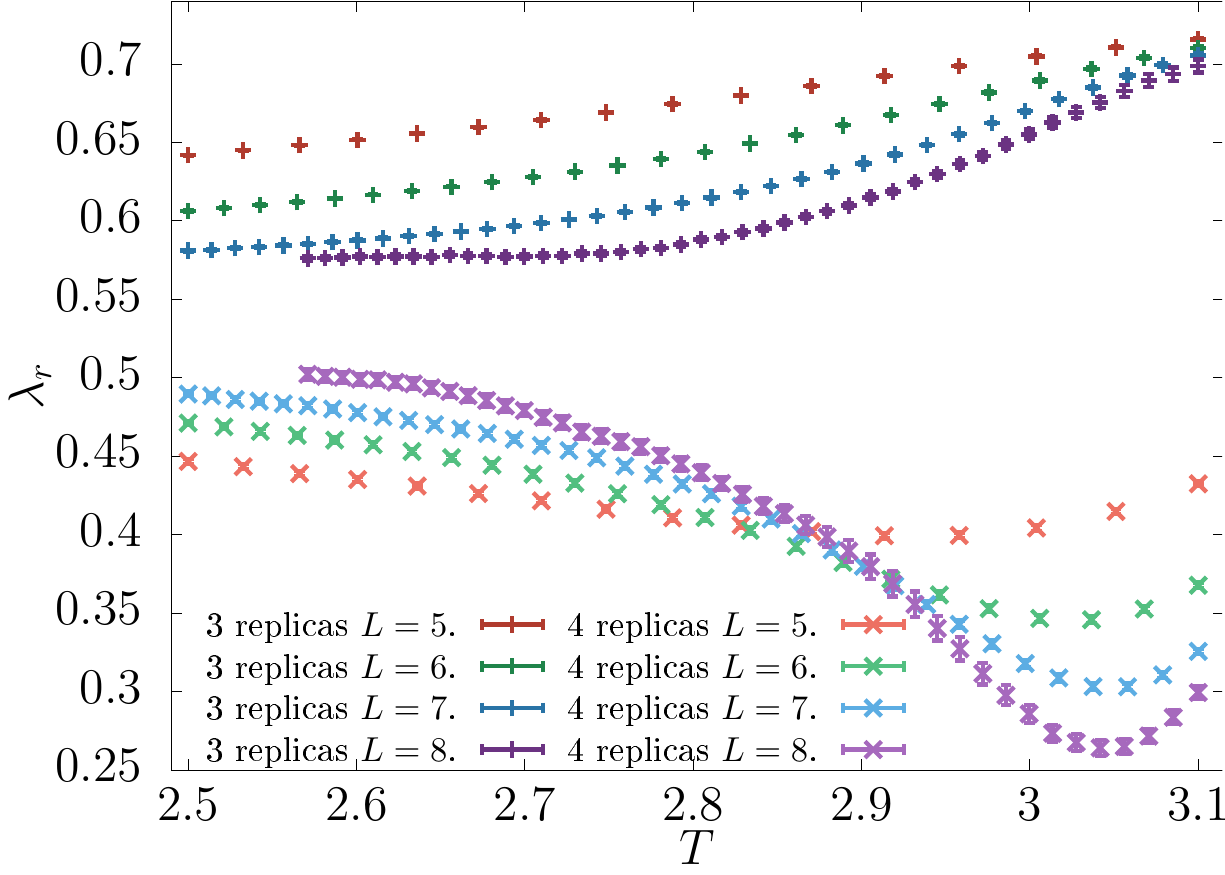}
    \caption{Three and four-replica estimators of $\lambda_r$, Eqs.~\eqref{eq:omega_3replicas} and \eqref{eq:omega_4replicas},  as a function of the temperature.  The six-replica estimator, shown in Appendix~\ref{app:lambda-def},  turned out to be compatible with ---but much noisier than--- the four-replica estimator.}
    \label{fig:lambda34}
\end{figure}

The validity of the prediction of $\lambda_r\in [0,1]$ can be addressed from Fig.~\ref{fig:lambda34} 
(we employ dark colors for the values of $\lambda_r$ computed
with three replicas and light colors for the four-replica estimator).  Notice that the scaling corrections of
the three- and four-replica estimators have opposite signs. Hence we
can assume that $\lambda_r(T_c^+)$ lies between the three- and
four-replica estimate of $\lambda_r$ for our largest lattice at our
estimate of the critical point. In this way, we conclude that $\lambda_r(T_c^+) = 0.52(6)$.

\section{Conclusions}\label{sec:conclusion} 
We have found  numerical evidences for a second order phase transition in the six-dimensional Ising spin glass in an external field by using state-of-the-art techniques for the Monte Carlo simulations and for the data analysis.
We have also performed a first computation of the critical exponents for this phase transition. Our finding  of a second-order phase transition in six dimensions in a field corroborates, and complements, the analysis based on high-temperature series expansions by Singh and Young~\cite{singh:17b}.

The difficulty in interpreting this finding lies in the absence of a stable fixed point in the one-loop studies for this problem based in the renormalization group by Bray and Roberts \cite{bray:80}, which opens the doors to nonperturbative scenarios. In fact, a recent study (conducted up to second order in perturbation theory) has found an additional and stable fixed point~\cite{charbonneau:17,charbonneau:19}. Furthermore, in Ref. \cite{Angelini:22} it has been claimed that the upper critical dimension is eight (rather than six). Our findings are most easily
interpreted in this context. An scenario in which a nonperturbative fixed point appears below eight dimensions, which controls the second order phase transition in six dimensions, seems quite natural.

However, we should stress that both techniques have their own limitations: truncation of the series in the case of the high-temperature expansions and the impossibility to simulate very large lattices in the case of numerical simulations. As a consequence, both techniques could miss  a crossover to a different behavior that would appear at higher orders in the series expansion, or for larger systems in the case of the simulations. This could be a crossover to a quasi-first-order transition scenario~\cite{holler:20}  or even to a no-phase transition scenario~\cite{mcmillan:84,fisher:86,bray:87,fisher:88,yeo:15}. If present, this change of behavior should occur at a crossover length larger than our simulated sizes $L=8$. We are unaware of quantitative estimates of this crossover length in the literature. We should stress, however, that in the absence of a field (when there is no doubt that six is the upper critical dimension) our simulated sizes $L\simeq 8$ are enough for a characterization of the phase transition ,including its associated logarithmic corrections; see Ref.~\cite{wang:93} and Appendix~\ref{H0_appendix}.

Let us conclude by stating that we consider that additional analytical and numerical studies are needed to fully understand the intriguing challenge of characterizing the critical behavior of a spin glass in a field in finite dimensions.

\begin{acknowledgments}
Our simulations have been carried out in the ADA cluster at the the \textit{Instituto de Computación Científica Avanzada de Extremadura} (ICCAEx), at Badajoz and at the Cierzo cluster at the \textit{ Instituto de Biocomputación y Física de Sistemas Complejos} (BIFI) in Zaragoza. We would like to thank both institutions and their staff.
This work was partially supported by Ministerio de Ciencia, Innovaci\'on y Universidades (Spain), Agencia Estatal de Investigaci\'on (AEI, Spain, 10.13039/501100011033), and European Regional Development Fund (ERDF, A way of making Europe) through Grants PID2020-112936GB-I00 and PID2022-136374NB-C21, by the Junta de Extremadura (Spain) and Fondo Europeo de Desarrollo Regional (FEDER, EU) through Grant No.\ IB20079. M.A.-J.was supported by Community of Madrid and Rey Juan Carlos University through Young Researchers program in R\&D (Grant CCASSE M2737). J.M.-G. was supported by the Ministerio de Universidades and the European Union ‘NextGenerationEU/PRTR’ through a 2021–2023 Margarita Salas grant.
\end{acknowledgments}
\appendix

\section{The Hamiltonian and the structure of the propagators} 
\label{appA}

The effective Hamiltonian describing the critical behavior of the $D$-dimensional Ising spin glass model in presence of a magnetic field $h$ can be written using the replica framework as~\cite{pimentel:02}
\begin{eqnarray} {\mathcal H} & = & \frac12 \int d^Dx
\left[ \frac12 \sum_{ab} (\nabla
  \phi_{ab})^2+ m_1 \sum_{ab} \phi_{ab}^2\right.  \nonumber \\ &+&m_2\sum_{abc}
  \phi_{ab}\phi_{ac}+ m_3\sum_{abcd} \phi_{ab} \phi_{cd}  \nonumber
  \\ & - &\left. \frac16 \widetilde{w}_1
  \sum_{abc} \phi_{ab} \phi_{bc} \phi_{ca}- \frac16 \widetilde{w}_2
  \sum_{ab} \phi_{ab}^3\right]\,,
\label{eq:Hamiltonian}
\end{eqnarray}
where the replicated overlap, $\phi_{ab}$, is a $n\times n$ symmetric matrix with zero in the diagonal and $n$ is the number of replicas ($n\to 0$).

The three fundamental modes of the correlation function are named replicon, anomalous and longitudinal modes. The anomalous and longitudinal modes become identical when one takes the limit of vanishing $n$. The fundamental propagators of the theory $ G_{\text{R}}(\boldsymbol{x}-\boldsymbol{y})$ and $ G_{\text{A}}(\boldsymbol{x}-\boldsymbol{y})$ are related to the natural propagators $G_1(\boldsymbol{x}-\boldsymbol{y})$, $G_2(\boldsymbol{x}-\boldsymbol{y})$, and $G_3(\boldsymbol{x}-\boldsymbol{y})$ as 
\begin{align}
\nonumber
    G_{\text{R}}(\boldsymbol{x}-\boldsymbol{y})&=G_1(\boldsymbol{x}-\boldsymbol{y})-2 G_2(\boldsymbol{x}-\boldsymbol{y})+ G_3(\boldsymbol{x}-\boldsymbol{y})\;,\\
\nonumber
    G_{\text{A}}(\boldsymbol{x}-\boldsymbol{y})&=G_{\text{L}}(\boldsymbol{x}-\boldsymbol{y})=G_1(\boldsymbol{x}-\boldsymbol{y})-4 G_2(\boldsymbol{x}-\boldsymbol{y})\\
    &+ 3 G_3(\boldsymbol{x}-\boldsymbol{y})\;,
\end{align}
where
\begin{align}
G_1(\boldsymbol{x}-\boldsymbol{y})&=  \overline{\langle s_{\boldsymbol{x}} s_{\boldsymbol{y}}\rangle^2}-q^2\,, \\
G_2(\boldsymbol{x}-\boldsymbol{y})&=  \overline{\langle s_{\boldsymbol{x}} s_{\boldsymbol{y}}\rangle \langle s_{\boldsymbol{x}} \rangle \langle s_{\boldsymbol{y}}\rangle}-q^2  \,, \\
G_3(\boldsymbol{x}-\boldsymbol{y})&= \overline{\langle s_{\boldsymbol{x}}\rangle^2 \langle s_{\boldsymbol{y}} \rangle^2}-q^2 \,.
\end{align}
with, as usual, $q=\overline{\langle s_{\boldsymbol{x}}\rangle^2}$ being the average overlap. 

The Fourier transform of  the correlation functions is
\begin{equation}
\label{eq:Fourier}
    \hat G_{\alpha}(\boldsymbol{k}) = \sum_{\boldsymbol{r}} e^{i\boldsymbol{k}\cdot \boldsymbol{r}} G_{\alpha}(\boldsymbol{r})\;\;,\; \alpha\in \{1,2,3,R,L,A\}\;.
\end{equation} 
Associated with each two-point correlation functions one can define a susceptibility $\chi_\alpha$ as 
\begin{equation}
    \chi_{\alpha}=\widehat{G}_{\alpha}(\boldsymbol{0})\; , \;\;\; \alpha\in \{1,2,3,\text{R},\text{L},\text{A}\}\;.
\end{equation}
In the particular case of $h=0$ case one finds $\chi_{\text{R}}=\chi_{\text{A}}=\chi_{\text{L}}$ (because $\chi_2=\chi_3=0$). However, as soon as $h\ne 0$, the replicon $\chi_{\text{R}}$ becomes significantly larger than $\chi_{\text{A}}=\chi_{\text{L}}$ in the spin glass phase. Details on how to compute the correlation functions with a multispin coding algorithm can be found in Appendix  \ref{app:details}.

\section{The Helmholtz and Gibbs free energies.}\label{app:Hemholtz-Gibbs}

With this Hamiltonian we can compute the associated free energy, the Helmholtz one, defined as
\begin{equation}
F(\lambda_{ab})=-\frac{1}{V} \ln \langle e^{\sum_{(ab)} V \lambda_{ab} \delta \tilde Q_{ab}}\rangle_r \,,
\end{equation}
where $(ab)$ denotes sum over $a\neq b$, the average over the replicated system is denoted as $\langle(\cdots)\rangle_r$ (see Ref.~\cite{parisi:13}) and 
\begin{equation}
    \delta \tilde Q_{a b} \equiv Q_{a b}-q\,,
    \label{eq:def_deltaq}
\end{equation}
with 
\begin{equation}
Q_{a b} \equiv \frac{1}{V} \sum_{\boldsymbol{x}} s_{\boldsymbol x}^a s_{\boldsymbol x}^b\,.
\end{equation}
This free energy allows us to compute the average value of the overlap
\begin{equation}
\langle \delta \tilde Q_{ab} \rangle_r =-\frac{\partial F}{\partial \lambda_{ab}}\,.
\end{equation}

The Gibbs free energy, $G(\delta Q_{ab})$, is just the Legendre transform of $F(\lambda_{ab})$ given by
\begin{equation}
G(\delta Q_{ab})=F(\lambda_{ab})+ \sum_{ab} \lambda_{ab} \delta Q_{ab}\,,
\end{equation}
$\lambda_{ab}$ being a function of $\delta Q_{ab}$
\begin{equation}
\lambda_{ab}  =\frac{\partial G}{\partial \delta Q_{ab}}\,.
\end{equation}

The Taylor expansion of the Helmholtz free energy (up to third order) is
\begin{eqnarray}
\nonumber
    F(\lambda)=&-&\frac{1}{2} \sum_{(ab)(cd)} G_{ab,cd} \lambda_{ab} \lambda_{cd}\\ 
    &-&\frac{1}{6} \sum_{(ab),(cd),(ef)} {\mathcal W}_{ab,cd,ef}  \lambda_{ab} \lambda_{cd} \lambda_{ef} \,.
\end{eqnarray}
The coefficients ${\cal W}_{ab,cd,ef}$ can take only eight different values in a RS phase, namely: 
${\mathcal W}_{ab,bc,ca}={\cal W}_1$,
${\mathcal W}_{ab,ab,ab}={\cal W}_2$,
${\mathcal W}_{ab,ab,ac}={\cal W}_3$,
${\mathcal W}_{ab,ab,cd}={\cal W}_4$,
${\mathcal W}_{ab,ac,bd}={\cal W}_5$,
${\mathcal W}_{ab,ac,ad}={\cal W}_6$,
${\mathcal W}_{ac,bc,de}={\cal W}_7$, and 
${\mathcal W}_{ab,cd,ef}={\cal W}_8$.

In terms of these ${\cal W}_i$ ($i=1,\dots, 8$) the cubic part of the Helmholtz free energy can be written as
\begin{eqnarray}
\nonumber
&&\sum_{(ab),(cd),(ef)} {\cal W}_{ab,cd,ef}  \lambda_{ab} \lambda_{cd} \lambda_{ef} \\
\nonumber
&=& \omega_1 \sum_{abc} \lambda_{ab} \lambda_{bc} \lambda_{ca}+\omega_2 \sum_{ab} \lambda_{ab}^3+\omega_3 \sum_{abc} \lambda_{ab}^2 \lambda_{ac}\\
\nonumber
&+&\omega_4 \sum_{abcd} \lambda_{ab}^2 \lambda_{cd} + \omega_5 \sum_{abcd} \lambda_{ab} \lambda_{ac} \lambda_{bd}+\omega_6 \sum_{abcd} \lambda_{ab} \lambda_{ac} \lambda_{ad} \\
&+& \omega_7 \sum_{abcde} \lambda_{ac} \lambda_{bc} \lambda_{de} + \omega_8 \sum_{abcdef} \lambda_{ab} \lambda_{cd} \lambda_{ef}
\,.
\end{eqnarray}

We quote here only the relationship between $\omega_1$ and $\omega_2$ (the two relevant terms in this context~\cite{parisi:13}) and the ${\cal W}_i$ \cite{temesvari:02b}:

\begin{equation}
\begin{aligned}
\omega_1 &= \mathcal{W}_1-3\mathcal{W}_5+3\mathcal{W}_7-\mathcal{W}_8\,,\\
\omega_2 &=\frac{1}{2} \mathcal{W}_2-3 \mathcal{W}_3+\frac{3}{2} \mathcal{W}_4+3 \mathcal{W}_5+2 \mathcal{W}_6-6 \mathcal{W}_7+2 \mathcal{W}_8 \,.
\label{eq:omegas}
\end{aligned} 
\end{equation}
that can expressed as:
\begin{equation}
\begin{aligned}
    \omega_1 &= \frac{1}{V}\sum_{\boldsymbol{x}\boldsymbol{y}\boldsymbol{z}} \overline{\langle s_{\boldsymbol{x}}
    s_{\boldsymbol{y}}\rangle_c \langle s_{\boldsymbol{y}} s_{\boldsymbol{z}}\rangle_c \langle s_{\boldsymbol{z}} 
    s_{\boldsymbol{x}}\rangle_c}\;,\\
    \omega_2 &= \frac{1}{2V}\sum_{\boldsymbol{x}\boldsymbol{y}\boldsymbol{z}}
    \overline{\langle s_{\boldsymbol{x}} s_{\boldsymbol{y}} s_{\boldsymbol{z}}\rangle^2_c}\;,
\end{aligned}
\end{equation}
where by $\langle(\cdots)\rangle_c$ we denote connected correlation functions. Notice that the coefficients of the terms of order $O(\lambda^m)$ in the expansion of the Helmholtz free energy, $F(\lambda)$, are the $m$-term connected susceptibilities.

At this point the Gibbs free energy can be written as 
\begin{equation}
\begin{aligned}
 G(\delta Q)&=\frac{1}{2} \sum_{(ab),(cd)} \delta Q_{ab} M_{ab,cd} \delta Q_{cd} \\
 &- \frac{w_1}{6} \sum_{abc} \delta Q_{ab} \delta Q_{bc} \delta Q_{ca} - \frac{w_2}{6} \sum_{ab} \delta Q_{ab}^3 \,.
\end{aligned}
\end{equation}

Observe that the cubic coupling in the Hamiltonian are written as $\widetilde{w}_1$, 
$\widetilde{w}_2$ and they are different from the coefficients $w_1$,$w_2$ of the Gibbs Free 
energy~\cite{parisi:13} (the so-called vertices in field theoretical language). 
The coefficients $\widetilde{w}_i$ and $w_i$ are, respectively, the bare and
dressed couplings, The bare and dressed couplings generally differ, they are only equal at the level of the tree
approximation in field theory.

\section{Details on the computation of the parameter $\lambda_r$ in numerical simulations}\label{app:lambda-def}

A good way to study the nature of the transition is by means of the parameter $\lambda_r$ defined as
\begin{equation}
 \lambda_r = \frac{w_{2,r}}{w_{1,r}}\;,  \label{eq:def_lambda}
\end{equation}
where $w_{1,r}$ and $w_{2,r}$ are the renormalized  exact vertices of the static replicated Gibbs free energy~\cite{parisi:13} defined in terms of the connected correlation functions computed at zero momenta as~\cite{parisi:88}
\begin{equation}
w_{i,r}=\frac{\omega_i}{\chi_{\text{R}} ^{3/2} \xi_2^{D/2}}\;\;,\; (i=1,2)\,.
\end{equation}
Therefore
\begin{equation}\label{eq:lambda_omegas}
\lambda_r=\frac{w_{2,r}}{w_{1,r}}=\frac{\omega_2}{\omega_1}\,.    
\end{equation}
In addition, it can be shown that the exact vertices ($w_i$) can be expressed in terms of the connected correlations functions at zero momenta ($\omega_i$) as~\cite{parisi:13}
\begin{equation}
  w_i=\frac{\omega_i}{\chi_{\text{R}}^3} \;\; (i=1,2)\,.
\end{equation}
and so
\begin{equation}\label{eq:lambda_omegas}
\lambda_r=\frac{w_{2,r}}{w_{1,r}}=\frac{\omega_2}{\omega_1}=\frac{w_2}{w_1}\,.    
\end{equation}
Notice that the parameter $\lambda_r$ is also related with the cubic couplings $\widetilde{w}_1$ and $\widetilde{w}_2$ of the Replica Symmetric Hamiltonian of Bray and Roberts \cite{bray:80} since, at the mean-field (MF) level, we have $\widetilde{w_i}=w_i$ $(i=1,2)$. Then, $\lambda_r<1$ signals the breaking point from RS to replica symmetry breaking~\cite{gross:85}.
Let us finally remark that all microscopic models, each of them with its own $\lambda_r$ parameter at a general temperature, will display  the same $T\to T_c$ limit for $\lambda_r=w_{2,r}/w_{1,r}$. Indeed, in Parisi's renormalization scheme both $w_{1,r}$ and $w_{2,r}$ are universal, i.e., independent of the microscopic Hamiltonian~\cite{parisi:88}.

In numerical simulations we compute $\omega_1$ and $\omega_2$ using Eq. \eqref{eq:omegas} with the 
$\mathcal{W}_i$ given by~\cite{parisi:13}
\begin{equation}
\begin{aligned}
&\mathcal{W}_1 \equiv V^2 \overline{\left\langle\delta \tilde Q_{12} \delta \tilde Q_{23} \delta \tilde Q_{31}\right\rangle}, \\
&\mathcal{W}_2 \equiv V^2 \overline{\left\langle \delta \tilde Q_{12}^3\right\rangle}, \\
&\mathcal{W}_3 \equiv V^2 \overline{\left\langle \delta \tilde Q_{12}^2 \delta \tilde Q_{13}\right\rangle}, \\
&\mathcal{W}_4 \equiv V^2 \overline{\left\langle \delta \tilde Q_{12}^2 \delta \tilde Q_{34}\right\rangle}, \\
&\mathcal{W}_5 \equiv V^2 \overline{\left\langle\delta \tilde Q_{12} \delta \tilde Q_{13} \delta \tilde Q_{24}\right\rangle}, \\
&\mathcal{W}_6 \equiv V^2 \overline{\left\langle\delta \tilde Q_{12} \delta \tilde Q_{13} \delta \tilde Q_{14}\right\rangle}, \\
&\mathcal{W}_7 \equiv V^2 \overline{\left\langle\delta \tilde Q_{12} \delta \tilde Q_{13} \delta \tilde Q_{45}\right\rangle}, \\
&\mathcal{W}_8 \equiv V^2 \overline{\left\langle\delta \tilde Q_{12} \delta \tilde Q_{34} \delta \tilde Q_{56}\right\rangle},
\end{aligned}
\label{eq:cubic_cumulants}
\end{equation}
and the overlap fluctuations $\delta \tilde Q_{ab}$ can be computed in terms of independent real replicas ($a$ and $b$) 
using Eq. (\ref{eq:def_deltaq}).

To compute each cubic cumulant $\mathcal{W}_i$ requires a number of different real replicas equal to the largest index in its expression. The RS field theory predicts that the amplitudes of the $\{\mathcal{W}_1,\dots,\mathcal{W}_8\}$ set are not independent \cite{parisi:13}. Using the linear relationship between them one can compute $\omega_1$ and $\omega_2$ in terms of three- and four-replica estimators. In particular, the three-replica estimators are~\cite{parisi:13}
\begin{equation}
    \begin{aligned}
        \omega_1^{(3)}\equiv \frac{11}{30}\mathcal{W}_1 + \frac{2}{15}\mathcal{W}_2\;,\\
        \omega_2^{(3)}\equiv \frac{4}{15}\mathcal{W}_1 -\frac{1}{15}\mathcal{W}_2\;,
    \end{aligned}
    \label{eq:omega_3replicas}
\end{equation}
and the four-replica ones~\cite{parisi:13}
\begin{equation}
\begin{aligned}
&\omega_1^{(4)} \equiv \frac{23 \mathcal{W}_1}{30}+\frac{\mathcal{W}_2}{20}-\frac{3 \mathcal{W}_3}{5}+\frac{9 \mathcal{W}_4}{20}-\frac{6 \mathcal{W}_5}{5}+\frac{\mathcal{W}_6}{2}\;, \\
&\omega_2^{(4)} \equiv \frac{7 \mathcal{W}_1}{15}+\frac{2 \mathcal{W}_2}{5}-\frac{9 \mathcal{W}_3}{5}+\frac{3 \mathcal{W}_4}{5}-\frac{3 \mathcal{W}_5}{5}+\mathcal{W}_6\;.
\end{aligned}
\label{eq:omega_4replicas}
\end{equation}
Within the framework of the RS theory, the values of the three- and four-replica estimators differ in general from the true values of $\omega_1$ and $\omega_2$ but coincide with them at the critical temperature~\cite{parisi:13}. This gives us the opportunity to check the validity of the RS theory by computing the thre-, four-, and six-replica estimators.

\subsection{$\mathbf{\lambda_r}$ infinite-temperature limit}
As a part of our analysis of the behavior of the observable $\lambda_r$, we have studied how far from the infinite-temperature limit our results are. In this section we show the computations of $\lambda_r$ in the infinite-temperature limit for three, four and six replicas.

Actually, the computation of $\lambda_r$ at infinite temperature is reduced to the computation of the cubic cumulants $\mathcal{W}_i$ with $i \in \{1,2,\dots,8\}$ in the same regime.

Let us start by computing the very simple case of the cumulant $\mathcal{W}_2$. Starting from the expression of the cumulant given by Eq.~\eqref{eq:cubic_cumulants} and Eq.~\eqref{eq:def_deltaq} we can develop the expression of $\mathcal{W}_2$:
\begin{equation}
    \begin{aligned}
        \mathcal{W}_2 & = V^2 \overline{\braket{\left(Q_{12}-q\right)^3}} = \\
        & = V^2 \overline{\braket{Q^3_{12}}} - 3 q \overline{\braket{Q^2_{12}}} + 2 V^2 q^3 \,.
    \end{aligned}
\end{equation}

In the infinite-temperature limit in which we are interested of, the value of $q$ tends to $0$, so the only relevant term is $V^2 \braket{Q^3_{12}}$. Expanding this term we get
\begin{equation}
    \braket{Q^3_{12}} = \dfrac{1}{V^3} \sum_{\boldsymbol{x}\boldsymbol{y}\boldsymbol{z}} \braket{s^1_{\boldsymbol{x}} s^2_{\boldsymbol{x}} s^1_{\boldsymbol{y}} s^2_{\boldsymbol{y}} s^1_{\boldsymbol{z}} s^2_{\boldsymbol{z}}} \, .
\end{equation}
In the infinite-temperature limit, the only relevant terms are those with $\boldsymbol{x}=\boldsymbol{y}=\boldsymbol{z}$. Other possibilities vanish when thermal averages are taken since thermal fluctuations are dominant in that regime. Thus, we have
\begin{equation}
    \braket{Q^3_{12}}_{T\to \infty} = \dfrac{1}{V^3} \sum_{\boldsymbol x} \braket{s^1_{\boldsymbol{x}} s^2_{\boldsymbol{x}} s^1_{\boldsymbol{x}} s^2_{\boldsymbol{x}} 
    s^1_{\boldsymbol{x}} s^2_{\boldsymbol{x}}} \, .
    \label{eq:cumulant_term_infinite_temperature}
\end{equation}
Spins appearing an odd number of times lead to cancellations in the Eq.~\eqref{eq:cumulant_term_infinite_temperature} when thermal averages are taken. On the contrary, spins appearing an even number of times can pair each other and contribute one unit to the sum.

This simple reasoning indicate us that, each time a replica is appearing an odd number of times in the expression of $\mathcal{W}_i$, that cubic cumulant will be $0$ in the infinite-temperature limit. That is the case for $\mathcal{W}_i$ with $i\geq 2$. However, for $\mathcal{W}_1$ we have
\begin{equation}
    \mathcal{W}_1 = V^2 \overline{\braket{Q_{12}Q_{23}Q_{31}}} - 3 q \overline{\braket{Q_{12}Q_{13}}} + 2V^2 q^3 \, .
    \label{eq:first_cubic_cumulant}
\end{equation}
In Eq.~\eqref{eq:first_cubic_cumulant}, the only nonzero term in the infinite limit, the first one, contains an even number of spins of each replica when expanding it. Therefore, this term will contribute with a factor $1$ when averages are taken. Then, $\mathcal{W}_1 = 1$ and $\mathcal{W}_i = 0$ for $i \geq 2$ in the infinite-temperature limit.

At this point, the computation of $\lambda_r$ is easy. From Eq.~\eqref{eq:def_lambda} we obtain
\begin{equation}
    \lambda_r=\dfrac{\omega_2}{\omega_1} = 0 \, .
\end{equation}
Similar computations are valid for the three- and four-replica cases. From Eqs.~\eqref{eq:omega_3replicas} and Eqs.~\eqref{eq:omega_4replicas} we have
\begin{equation}
    \lambda_r^{(3)} = \dfrac{\omega^{(3)}_2}{\omega^{(3)}_1} = \dfrac{8}{11} \, ,
\end{equation}
\begin{equation}
    \lambda_r^{(4)} = \dfrac{\omega^{(4)}_2}{\omega^{(4)}_1} = \dfrac{14}{23} \, .
\end{equation}

\section{Numerical simulation details} \label{app:details}
To study the six dimensional Ising spin glass in the presence of a magnetic field we have written several computer programs in \texttt{C} language. In our simulations, we use an offline analysis approach, meaning our simulation program writes configurations of spins and then a different analysis program computes the correlation lengths and other magnitudes of interest. Finally, a third program is in charge of the statistical analysis of the data, using  the Jackknife technique \cite{Jacknife_PY, efron1982jackknife}. It is worth mentioning that both, the simulation and the analysis program, make use of the POSIX thread libraries, which allow us to create different threads that run parallel on different cores. 

We have simulated the Edwards-Anderson model in a six dimensional hypercube with periodic boundary conditions, using a multispin coding Monte Carlo simulation and performing a parallel tempering  \cite{hukushima:96,marinari:98b} proposal every 20 Monte Carlo sweeps (MCS). For each sample, we have simulated six replicas. The sizes of the side $L$ of the hypercube range from $L=5$ to $L=8$. The value of the magnetic field $h$ has been set to $h=0.075$. The number of samples used to obtain the results that we present can be consulted in Table \ref{tabla_sim}, along with other information about the simulations. 

\begin{table}[h]
\centering
\begin{tabular}{c c c c c c} \hline 
$L$ & ~~No. samples~~~~& No. temp. & ~$T_\mathrm{min}$ & ~$T_\mathrm{max}$ & ~~~~ps/spin flip               \\ \hline \hline
5            & 25600              & 16                  & 2.50    &3.10          & 16                                     \\ 
6            & 25600              & 24                   & 2.50    &3.10         &  12                                     \\ 
7            & 25600              & 36                    & 2.50    &3.10         &  20                                      \\ 
8            & 5120                & 44                    & 2.57    &3.10        &  16                                       \\ \hline \hline
\end{tabular}
\caption{Some parameters of our simulation. The first column refers to the size of the side of the hypercube. In the second column we present the number of samples analyzed. The third column shows the number of temperatures simulated for each size. This number has been chosen in a way that ensures the random walk in temperatures to be sufficiently ergodic. The fourth and fifth columns refer to the lower and upper values of the temperature interval. Finally, the last column shows the number of picoseconds (ps)  per spin flip of the simulation.  }
\label{tabla_sim}
\end{table}
To ensure that we measure the equilibrium properties of the system, the thermalization of each sample must be studied individually. We have designed a thermalization protocol based on Ref. \cite{billoire:18} that works in the following way.

First we simulate our supersample (a package of 128 samples) during a sufficiently large number of MCS for most of the samples to be thermalized. This value $\mathcal{N}_\mathrm{sim}$ was determined by preliminary runs.

During the simulation, the random walk in temperatures induced by the parallel tempering algorithm for each of the 128 samples is registered. Then, when the simulation ends, we study this random walk, and compute from it the integrated autocorrelation time $\tau_{\mathrm{int}, f}$ for several magnitudes $f$, related with the random walks \cite{billoire:18}. We take the largest of these integrated autocorrelation times, $\tau_{\mathrm{int}, f*}$, and make the assumption $\tau_{\mathrm{int}, f*}\sim \tau_\mathrm{exp}$. The value of the exponential autocorrelation time enables us to check if a given sample is thermalized. In particular, we consider a sample to be thermalized if its simulation time $\tau_\mathrm{sim}$ is thirty times bigger than  $ \tau_\mathrm{exp}$  (i.e $\mathcal N_\mathrm{sim}> 30  \tau_\mathrm{exp}$).

When in a supersample  a nonequilibrated sample is found, we proceed as follows. The last configurations of those samples which are not thermalized are extracted from the corresponding supersample. Then a new supersample is built grouping nonthermalized samples and their simulation time is doubled. Finally, the thermalization check is repeated, $\tau_\mathrm{exp}$ is measured, and if the criteria is not fulfilled, the simulations are extended once again. This process is repeated until all samples reach their equilibrium states. At the end, the samples are reintroduced into their original supersamples, ready to be analyzed.

\section{Details of multispin coding algorithms} \label{Multi-spin-Apendice}
One of the most complex tasks we have faced when trying to generate the results that we present here is the elaboration of coding algorithms. In this appendix we show some examples that can give a taste of how the problem is approached. 
\subsection{Metropolis algorithm}
Current CPUs are able to execute one-clock-cycle instructions over registers (or words) of 128 bits. The 128-bit words are coded by using the Intel Intrinsics'\footnote{\href{https://www.intel.com/content/www/us/en/docs/intrinsics-guide/index.html}{https://www.intel.com/content/www/us/en/docs/intrinsics-guide/index.html}} variables $\_\,\_$\texttt{m128i} consisting in 128 Boolean variables. We benefit from this 128-bits words by simulating at once 128 samples. 

Let us define a vector \texttt{S[V]} of type $\_\,\_$\texttt{m128i}, where \texttt{V} is the number of spins of our system. Each element of the vector \texttt{S[i]}, contains 128 Boolean variables representing the value of the spin at the $i$-th position for the 128 samples. Analogously, we can define a vector \texttt{J[6*V]} in which each element \texttt{J[i]} contains the value of the coupling for the 128 samples at the $i$-th position. Now, the problem reduces to code the Metropolis algorithm in bitwise operations to improve the performance up to 128 times. We work with the following assignment 
 \begin{equation}
     s=+1\to b_s=+1, \;\;\text{and}\;\; s=-1\to b_s=0\;,
 \end{equation}
 \begin{equation}
     J=+1\to b_J=0 \;\; \text{and} \;\; J=-1\to b_J=1\;,
 \end{equation}
 where the $b$ variables are the bit variable of our program. This seemingly arbitrary election has a virtue: the equivalence 
 \begin{equation}\label{prodxor}
     \texttt{b}_{s1} \wedge \texttt{b}_J \wedge \texttt{b}_{s2} = (1-Js_1s_2)/2\,,
 \end{equation}
 where $\wedge$ represents the Boolean \textit{exclusive or} (XOR).
 
We use a variation from the original Metropolis algorithm to decide if a given spin is flipped or not. The Hamiltonian is composed by an interaction term $\mathcal{H}_J$ and a magnetic field term $\mathcal{H}_h$. The probability of performing a given spin flip is determined by the product
\begin{equation}
    \text{min}\{1, \exp(-\Delta\mathcal{H}_J)\} \text{min}\{1, \exp(-\Delta\mathcal{H}_h)\} \;.
\end{equation}
This election is not completely equivalent to the standard Metropolis algorithm, but it verifies the detailed balance condition and has the virtue of simplifying the implementation.  One can execute this spin flip simultaneously for the 128 samples having an $\_\,\_$\texttt{m128i} variable \texttt{flip} which encodes the information. If the spin from the $i$-th sample must be flipped, then $\texttt{flip}^i = 1$, if it must remain in the same position  $\texttt{flip}^i = 0$. Then one can carry out the flip decision simultaneously using the \texttt{XOR} operator 
\begin{equation}
    \texttt{S[site]} =\texttt{S[site]} \wedge \texttt{flip[site]} \;.
\end{equation}
The \texttt{flip} variable has two different contributions, one coming from the difference in the interaction energy considering the spin neighbors, \texttt{flipJ}, and other coming from the spin alignment with the field $h$, \texttt{fliph}, so that 
\begin{equation}
    \texttt{flip} = \texttt{flipJ} \;\& \;\texttt{fliph}\;.
\end{equation}

Since we only consider nearest-neighbour interactions this energy difference coming from the interactions with neighboring sites will be 
\begin{equation}
    \Delta \mathcal{H}_J = - 2s_{\boldsymbol{x}}\sum_{\boldsymbol{y}} J_{\boldsymbol{x} \boldsymbol{y}} s_{\boldsymbol{y}}\;,
\end{equation}
where the sum is restricted over the $12$ neighbours of $s_{\boldsymbol{x}}$. In the $\{+1,-1\}$ base the energy difference $\Delta \mathcal{H}_J$ can take 12 values ranging from $-24$ to $+24$ in steps of 4. To compute $\Delta E_J$ in the bit base $\{0,1\}$ we define $n_\mathrm{un}$, which represent the number of unsatisfied couplings of a given spin. The maximum possible value of $n_\mathrm{un}$ is 12, and the energy difference is $\Delta \mathcal{H}_J=24-2n_\mathrm{un}$. To compute $n_\mathrm{un}$ we select a given spin \texttt{S[site]} and check if a given coupling is satisfied using  Eq. \eqref{prodxor}. This information is then encoded in  eight 128-bit variables. To account for the heat bath effect, we generate a random number $R\in[0,1)$ and check if $R<\exp(-\beta\Delta \mathcal{H}_J)$ for each sample. We use the same random number for all samples, so we can check if it surpass a barrier of $\exp(-4\beta)$, $\exp(-8\beta)$, $\dots$, $\exp(-24\beta)$. The number of barriers surpassed are encoded in three new \texttt{$\_\,\_$\texttt{m128i}} variables named \texttt{id1}, \texttt{id2}, \texttt{id3}. We assign \texttt{flipJ} a value of one if 
\begin{equation}
    \texttt{id1}+ 2 \texttt{id2} + 4 \texttt{id3} + n_\mathrm{un} \ge 6 
\end{equation}
  Computationally, taking this decision takes a total of 54 Boolean operations. 

Finally one check if after the flip the spin is aligned with the field $h$, and if so make the assignment $\texttt{fliph}=1$. 
This process is repeated for every spin of the system. 

\subsection{Multispin coding for correlation functions}\label{MSC_corr}
Our goal is to compute the correlation functions $G_{\text{R}}(x)$ and $G_{\text{A}}(x)$ as defined in section \ref{appA}. For convenience, we define two new propagators $\Gamma_1$ and $\Gamma_2$ as 
\begin{equation}
    \Gamma_1(\boldsymbol{x}-\boldsymbol{y}) = \overline{\langle(s_{\boldsymbol{x}}-\langle s_{\boldsymbol{x}}\rangle) (s_{\boldsymbol{y}}-\langle s_{\boldsymbol{y}}\rangle) \rangle ^2 }\; ,
\end{equation}
\begin{equation}
    \Gamma_2(\boldsymbol{x}-\boldsymbol{y}) = 
    \overline{\langle s_{\boldsymbol{x}} s_{\boldsymbol{y}}\rangle^2} - \overline{\langle s_{\boldsymbol{x}} \rangle^2  \langle s_{\boldsymbol{y}} \rangle^2}\;. 
\end{equation}
It can be easily shown that 
\begin{equation}
     G_{\text{R}}(\boldsymbol{x}-\boldsymbol{y})=\Gamma_1(\boldsymbol{x}-\boldsymbol{y})\; , 
\end{equation}
and 
\begin{equation}
     G_{\text{A}}(\boldsymbol{x}-\boldsymbol{y})=2\Gamma_1(\boldsymbol{x}-\boldsymbol{y})-\Gamma_2(\boldsymbol{x}-\boldsymbol{y})\;,
\end{equation}
so it is enough for us to compute $\Gamma_1$ and $\Gamma_2$ in order to obtain the replicon and anomalous propagators. They can be computed with the help of replicas as 
\begin{equation}
    \Gamma_1(\boldsymbol{x}-\boldsymbol{y}) = \overline{\langle \varphi_{ab;cd}(\boldsymbol{x})\varphi_
    {ab;cd}(\boldsymbol{y})\rangle}\;,
\end{equation}
\begin{equation}
    \Gamma_2(\boldsymbol{x}-\boldsymbol{y})= \overline{\langle \Delta_{ab;cd}(\boldsymbol{x})\Delta_{ab;cd}(\boldsymbol{y})\rangle}\;,
\end{equation}
where 
\begin{equation}\label{def_phi}
    \varphi_{ab;cd}(\boldsymbol{x}) = \frac{(s_{\boldsymbol{x}}^a-s_{\boldsymbol{x}}^b)(s_{\boldsymbol{x}}^c-s_{\boldsymbol{x}}^d)}{2}\;,
\end{equation}
\begin{equation}\label{def_Delta}
    \Delta_{ab;cd}(\boldsymbol{x}) = \frac{(q_{\boldsymbol{x}}^{ab}-q_{\boldsymbol{x}}^{cd})}{\sqrt{2}}\;.
\end{equation}
Notice that one need at least four replicas to compute the two points connected correlation functions estimators. We would like to evaluate the fields  $\varphi_{ab;cd}(x)$ and $\Delta_{ab;cd}(x)$ simultaneously for the 128 systems, so we are forced to do it just using Boolean operations. One can easily check that  $\varphi_{ab;cd}(x)/2$ can take three values: $+1, 0, -1$. Then, its value can be completely determine using three auxiliary Boolean variables defined as 
\begin{equation}
    \varphi_1 = (s^a\wedge s^b)\&(s^c\wedge s^d)\;,
\end{equation}
\begin{equation}
    \varphi_2 = (s^a\wedge s^d)\;,
\end{equation}
\begin{equation}
    \varphi_3 =  \varphi_1  \& \varphi_2 \;.
\end{equation}
In particular, $\varphi_1=1$ if the field in this point  $\varphi_{ab;cd}(x)/2$ is not zero and $\varphi_3=1$ if it is $+1$. Then one can compute the value of the field in a given hyperplane $P$ using a Boolean sum function and computing 
\begin{equation}
   \frac{1}{2} \sum_{x\in P} \varphi_{ab;cd}(x)= 2\sum_{x\in P}\varphi_3(x)- \sum_{x\in P}\varphi_1(x)\;.
\end{equation}
To compute $\Delta_{ab;cd}(x)$ one proceeds in a similar way. In particular, the auxiliary Boolean variables are 
\begin{equation}
    \Delta_1 = (s^a\wedge s^b)\wedge (s^c\wedge s^d) \;,
\end{equation}
\begin{equation}
      \Delta_2 =  \Delta_1  \& (s^c\wedge s^d)\;.
\end{equation}
The value of $\Delta_1(x)$ tells you if $\Delta_{ab;cd}(x)/\sqrt{2}$ is +1 and $\Delta_2$ if it is not zero.

\section{Computing the critical exponents}\label{app:quotients}
The value of the critical exponents can be estimated using the quotient method. Usually, the exponent $\eta$ would be extracted from the replicon susceptibility $\chi_{R}$, but to avoid the problematic $\boldsymbol{k}=\boldsymbol{0}$ wave vector we compute it from $\mathcal{F}$ as defined in Eq. (14), which has the same scaling behavior of $\chi_{R}$. The quotient method idea is to compute
\begin{equation}
    \mathcal{Q}_{\mathcal{F}}= \frac{\mathcal{F}^{L_2}(T_c)}{\mathcal{F}^{L_1}(T_c)} =  \left(\frac{L_2}{L_1}\right)^{2-\eta} + \dots\;\;, 
\end{equation}
and then, to leading order,
\begin{equation}\label{eq:eta}
    \eta= 2-\frac{\ln \mathcal{Q}_{\mathcal{F}}}{\ln(L_2/L_1)}\;.
\end{equation}
From our data we can obtain an estimation of $\eta$, given the three combinations of increasing values $L_1$ and $L_2$. 

The $\nu$ exponent can also be computed by using the quotient method. A standard way to do it would be studying the correlation length derivative with respect to the inverse temperature $\beta$. However, once again we avoid the $\boldsymbol{k}=\boldsymbol{0}$ mode and compute $\nu$ by means of the derivative of the $R_{12}$, defined in Eq. (12), as 
\begin{equation}
    \mathcal{Q}_{\partial_\beta R}= \frac{\partial_\beta R_{12}^{L_2}(T_c)}
    {\partial_\beta R_{12}^{L_1}(T_c)} = \left(\frac{L_2}{L_1}\right)^{1/\nu} + \dots \;,    
\end{equation}
\begin{equation}
    \nu = \frac{\ln(L_2/L_1)}{\mathcal{Q}_{\partial_\beta R}}\;. 
\end{equation}
The values of $\partial_\beta R_{12}$ have been obtained by deriving a polynomial fit of the sixth degree of $R_{12}$.

\section{The $h=0$ transition }\label{H0_appendix}
\subsection{Theory}
As a previous step to the analysis of our system, we have studied the six-dimensional model of an Ising spin glass without a magnetic field, described by the Hamiltonian 
\begin{equation}\label{hamiltonian_H0}
    \mathcal{H} = - \sum_{\langle \boldsymbol{x}, \boldsymbol{y}\rangle} J_{\boldsymbol{x}\boldsymbol{y}}s_{\boldsymbol{x}} s_{\boldsymbol{y}}\;,
\end{equation}
where  $J_{\boldsymbol{x}\boldsymbol{y}}=\pm 1$ with equal probabilities, $s_{\boldsymbol{x}}$ are Ising spin variables and the sum is restricted over nearest neighbours. The system has been studied by means of standard numerical Metropolis Monte Carlo simulations (with a nonmultispin coding program) and the help of the parallel tempering algorithm. We check the existence of a phase transition by computing the Binder cummulant,  defined as 
\begin{equation} \label{Binder}
g = 1- \frac{\overline{\langle q^4_J\rangle}}{3 \overline{\langle q^2_J \rangle}^2}\;,
\end{equation}
where the overlap computed in a given sample, $q_J$, is 
\begin{equation}
    q_J = \frac{1}{V}\sum_{\boldsymbol x} s_{\boldsymbol{x}} \tau_{\boldsymbol{x}}\;, 
\end{equation}
where $\{s_{\boldsymbol{x}}\}$ and $\{\tau_{\boldsymbol{x}}\}$ are, as usual, two replicas of the system evolving with the same disorder but different initial conditions.
The correlation length $\xi_2$ is defined using  the two point (nonconnected) correlation function
\begin{equation}
     G(\boldsymbol{x}-\boldsymbol{y}) = \overline{\langle s_{\boldsymbol{x}}s_{\boldsymbol{y}} \rangle^2}\;.
\end{equation}
We check the values of the critical exponents and, since we are at  the upper critical dimension $D_U$, we also check the values of the logarithmic corrections. 

In the presence of logarithmic corrections, the scaling laws near the critical point for the correlation length and the susceptibility  must be modified as \cite{Amit-Martin,kenna:06} 
\begin{equation}
    \xi_2 \sim |t|^{-\nu} |\ln |t||^{\hat \nu}\,,
\end{equation}
\begin{equation}
    \chi_{SG} \sim |t|^{-\gamma}|\ln |t||^{\hat{\gamma}}\,.
\end{equation}

At the infinite volume critical point, the correlation length defined on a box of size $L$ behaves as \cite{kenna:06}
\begin{equation}
    \xi_2 \sim L(\ln L)^{\hat{q}}\;,
\end{equation}
and the susceptibility
\begin{equation}
    \chi_{SG} \sim L^{2-\eta} (\ln L)^{\hat{a}}\;,
\end{equation}
so the complete scaling behavior for the six dimensional spin glass of $\xi/L$ and $\chi$ is \cite{ruiz-lorenzo:17}
\begin{equation} \label{scaling_corr_xi}
    \xi_2/L = |\ln L|^{\hat{q}} f_{\xi}\left( L^{2} t / (\ln L )^{4/3} \right)\,,
\end{equation}
and
\begin{equation}
    \chi_{SG} = L^{2} |\ln L |^{\hat{a}}  f_{\chi}\left( L^{2} t / (\ln L )^{4/3} \right)\,,
    \label{scaling_corr_chi}
  \end{equation}  
  where we have used the MF values $\nu=1/2$ and $\eta=0$.
The logarithmic corrections have been computed making use of the renormalization group and their values are \cite{ruiz-lorenzo:98, kenna:06,ruiz-lorenzo:17}: $\hat{q} = 1/6$ and $ \hat{a} = 2/3$.

\subsection{ Simulation results}
We have carried out simulations for systems with size $L$ ranging from $L=3$ to $L=9$ and for temperatures between $T_\mathrm{min}=3.012$ and $T_\mathrm{max}=3.031$. 
The number of MCS used for each sample as well as the number of samples are reported  in Table \ref{tabla:datos_h0}.

\begin{figure}
    \centering
    \includegraphics[scale=0.65]{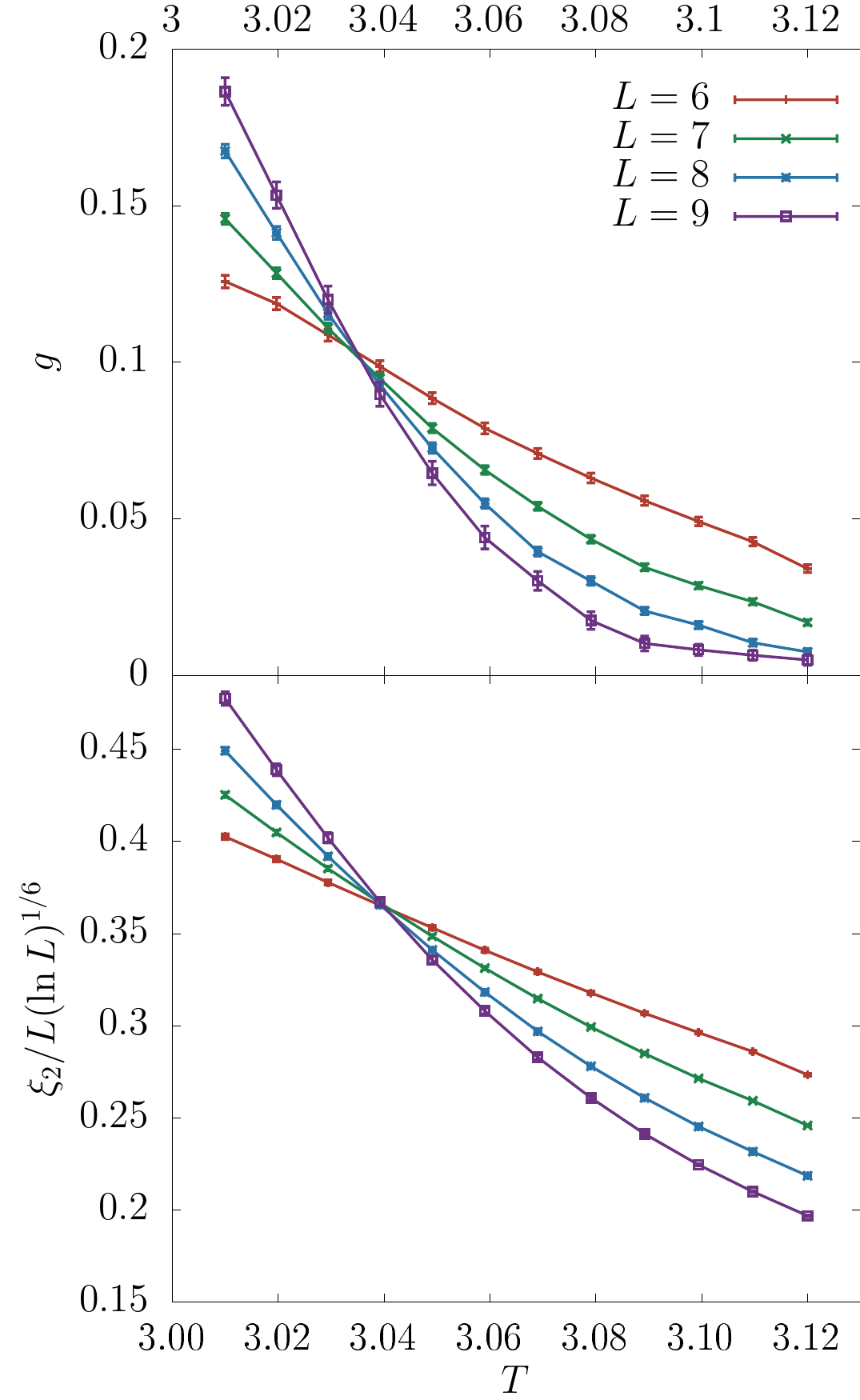}
    \caption{Binder cumulant $g$ (on the \textbf{top}), Eq. \eqref{Binder}, and the second moment correlation length $\xi_2$ divided by the lattice size $L$ and by the logarithmic correction term $(\ln L)^{1/6}$ (on the \textbf{bottom}) as a function of temperature for lattice sizes between $L=6$ and $L=9$. The presence of an intersection between the curves at  temperatures $T_c^{B}$ and $T_c^{\xi}$ signals the existence of a continuous phase transition.}
    \label{fig:double_tc}
\end{figure}

\begin{table}[h]
\begin{tabular}{ c c c } \hline
\hline
$L$ ~~& ~~~MCS~~~ & ~~~\#Samples \\ \hline
3            & 409600       & 8500             \\ \hline
4            & 409600       & 6500             \\ \hline
5            & 409600       & 5500             \\ \hline
6            & 409600       & 4000             \\ \hline
7            & 409600       & 4000             \\ \hline
8            & 204800       & 3000             \\ \hline
9            & 307200       & 1000             \\ \hline  \hline
\end{tabular}
\caption{Table showing the number of samples [independent systems with different values of the interaction coupling J's, recall Eq.~(1)] simulated and the number of MCS made during each simulation for each value of the lattice size $L$. }
\label{tabla:datos_h0}
\end{table}
The critical temperature have been computed by studying the Binder cumulant and the correlation length (see  Fig. \ref{fig:double_tc}) and we have obtained a value of 
\begin{equation}
    T_c=3.033(1)\,,
\end{equation}
which is compatible with the previous known results obtained via simulations $T_c=3.035(10)$~\cite{wang:93} and a high temperature expansion $T_c=3.027(5)$~\cite{klein:91}. 

\begin{figure}[h]
    \centering
    \includegraphics[scale=0.65]{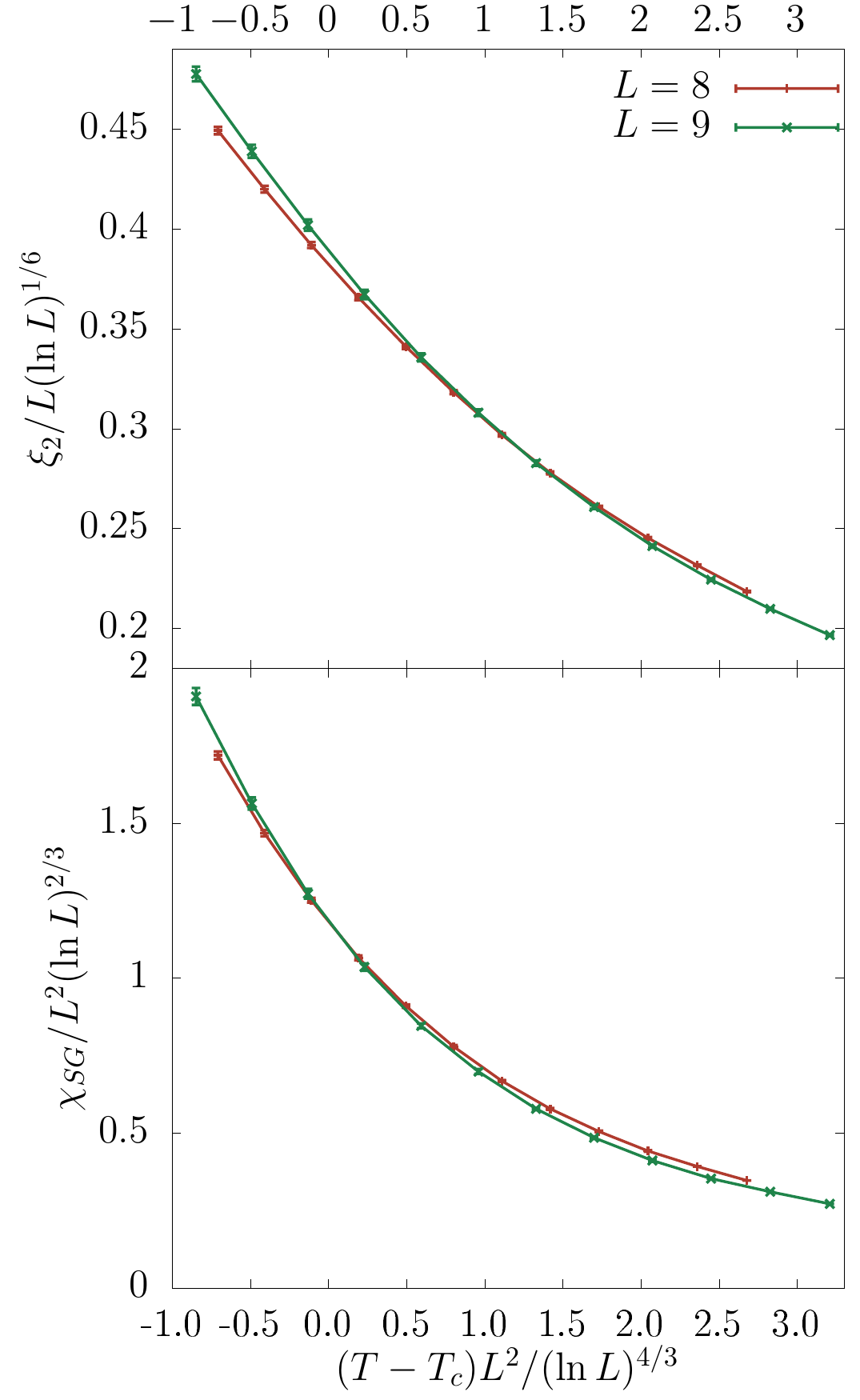}
    \caption{Second moment correlation length $\xi_2$ divided by the lattice size $L$ and by the logarithmic correction term $(\ln L)^{1/6}$ (on the \textbf{top}) and the spin glass susceptibility $\chi_{SG}$ divided by  $L^{2-\eta}$ (using the mean field result $\eta=0$) and by the logarithmic correction term $(\ln L)^{2/3}$ (on the \textbf{bottom}) as a function of the argument of the scaling functions $f_{\xi}$ and $f_{\chi}$ for lattice sizes $L=8$ and $L=9$. Since Eqs. \eqref{scaling_corr_xi} and  \eqref{scaling_corr_chi}  hold asymptotically for large vales of $L$ we would expect the two curves to collapse into a single one only if the used values of the logarithmic correction critical exponents $\hat{q}$ and $\hat{a}$ are the correct one.}
    \label{fig:double_colapso}
\end{figure}

The theoretical value of the critical exponents and logarithmic corrections can be checked using the collapse method, i.e., by representing $\xi_2/L |\ln L|^{1/6}$ and $\chi_{SG}/ L^{2-\eta} |\ln L |^{2/3} $ vs $ L^{1/\nu} t / (\ln L )^{4/3}$. Since the expressions \eqref{scaling_corr_xi} and \eqref{scaling_corr_chi} holds asymptotically for large values of $L$, one would expect the curves for different lattice size to collapse into a single curve near the critical temperature when $L \gg 1 $ if and only if the theoretical values of $\nu$, $\eta$,  $\hat{q}$ and $\hat{a}$ are the correct ones. The results can be checked in Fig. \ref{fig:double_colapso}.

\begin{figure}[!h]
    \centering
    \includegraphics[scale=0.65]{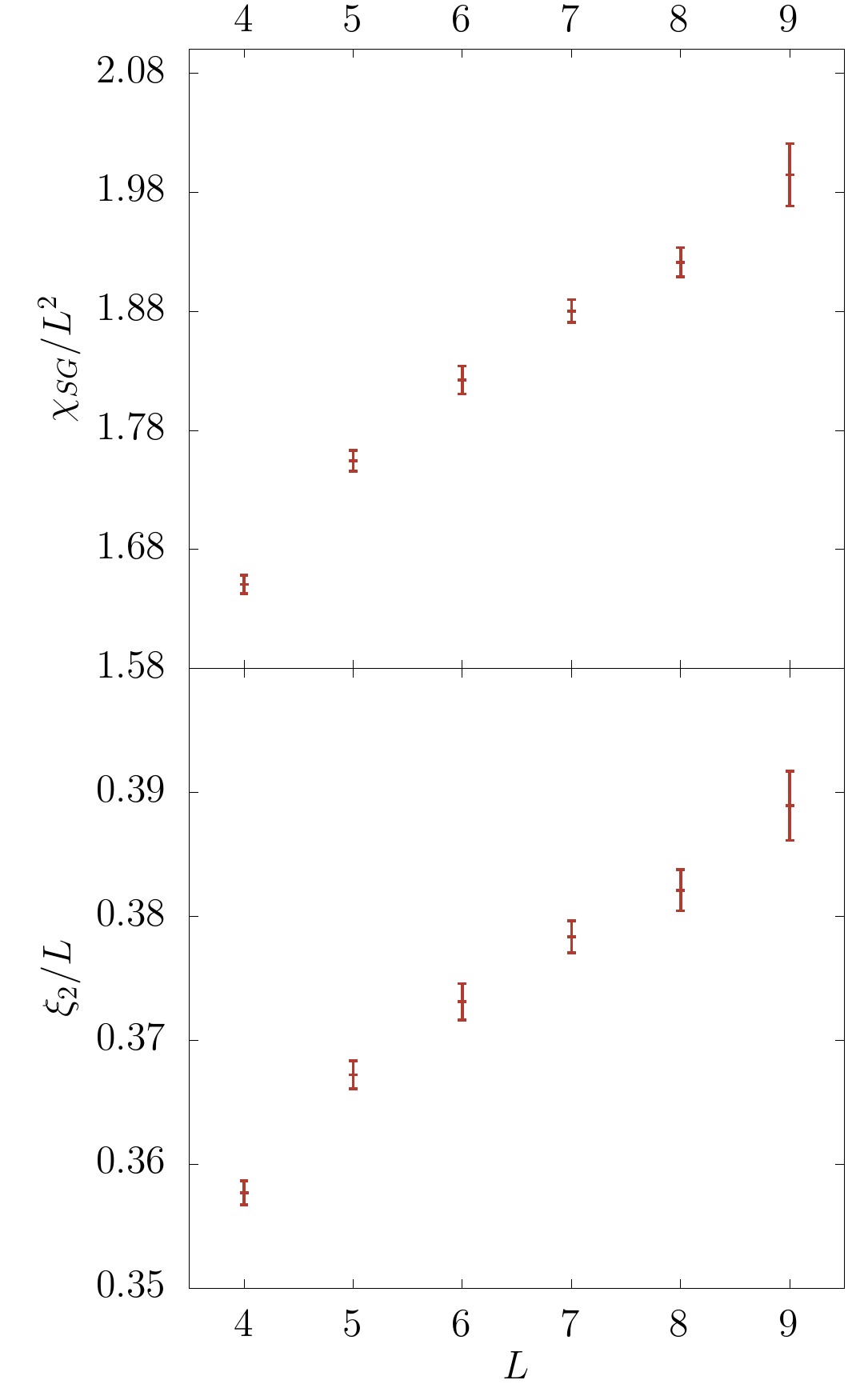}
    \caption{Plot of $\chi_{SG}(T_c)/L^2$ (on the \textbf{top}) and  $\xi_2 (T_c)/L$ (on the \textbf{bottom}) vs the lattice size $L$. The data shows a monotonic growth, a strong indication of the presence of logarithmic corrections.}
    \label{fig:no_correciones}
\end{figure}

Another strong indication of the presence of logarithmic corrections is presented in Fig. \ref{fig:no_correciones} in which we plot 
 $\chi_{SG}(T_c)/L^2$ and  $\xi_2 (T_c)/L$, both  showing a clear growing with the lattice size. This behavior was already observed in Ref. \cite{wang:93}  for the spin-glass susceptibility and enabled them to estimate the value of the logarithmic correction's power, which was previously unknown at that time. 
 
\begin{figure}[!h]
    \centering
    \includegraphics[scale=0.65]{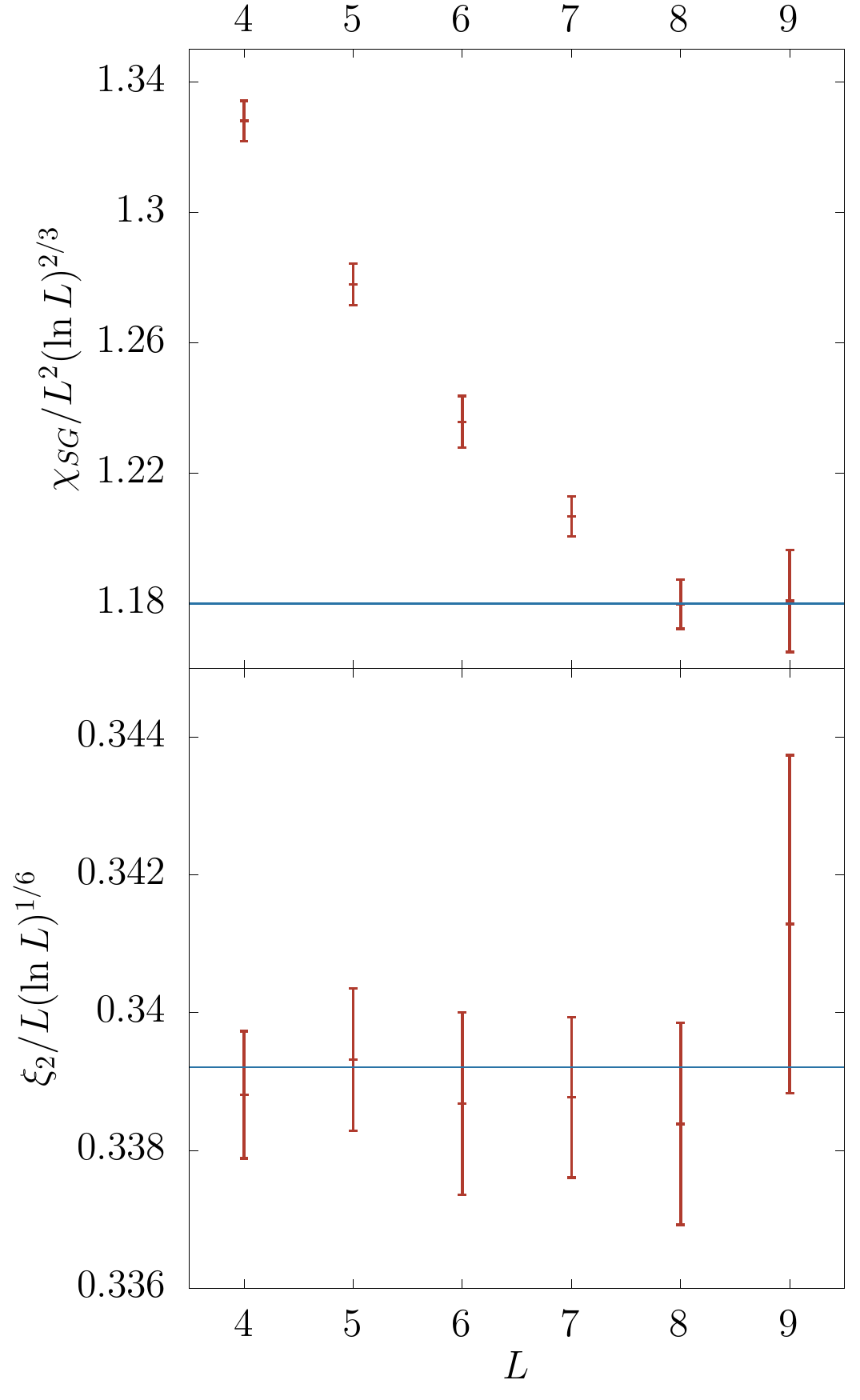}
    \caption{Plot of $\chi_{SG}(T_c)/[L^2 (\ln L)^{2/3}]$ (on the \textbf{top}) and  $\xi_2 (T_c)/[L (\ln L) ^{1/6}]$ (on the \textbf{bottom}) vs the lattice size $L$. Our data approach asymptotically to a constant value for $L\gg1$, from which we conclude that they are compatible with the values $\hat{a}=2/3$ and  $\hat{q} = 1/6$. The constant behavior for the bottom plot was expected since the scaling of $\xi$ was used to determine the value of $T_c$.}
    \label{fig:double_correciones}
\end{figure}

However, it is difficult to discern with precision the correct value of the logarithmic correction from Fig. \ref{fig:double_colapso}. Despite this, we can check the outcome of the analytical computation, $\hat{q}$ and $\hat{a}$, by representing $\chi_{SG}(T_c)/[L^2(\ln L)^{2/3}]$ and $\xi_2 (T_c)/[L(\ln L)^{1/6}]$ vs the lattice size $L$. If the scaling behavior \eqref{scaling_corr_xi} and \eqref{scaling_corr_chi} holds, and the theoretical values of the critical exponents are compatible with our data, then we would observe a constant behavior (represented by the blue line in Fig. \ref{fig:double_correciones}).

From the analysis above, we conclude that our simulation results are compatible with the MF values of the critical exponents and with the theoretical values of the logarithmic corrections $\hat{q} = 1/6$ and $ \hat{a} = 2/3$.

\section{$\Lambda$ cumulants}\label{app:Lambda}

One could have addressed the computation of the two renormalized constants $w_{i,r}$ ($i=1,2$), however, they presents strong finite-size effects. Instead, we have resorted to the computation of the $\Lambda_i$ cumulants introduced in Ref.~\cite{fernandez:22}
\begin{equation}\label{eq:Lambda_cummulant}
    \Lambda_i = \dfrac{\omega_i}{\chi^{3/2}_{\text{R}} L^{D/2}} \,\,\, \mathrm{with} \,\, i \in \{1,2\} \, .
\end{equation}
The $\Lambda_i$ cumulants scales in the same way as a dimensionless observable, like the Binder cumulant or $\xi_2/L$.

\begin{figure}[h]
    \centering
    \includegraphics[scale=0.75]{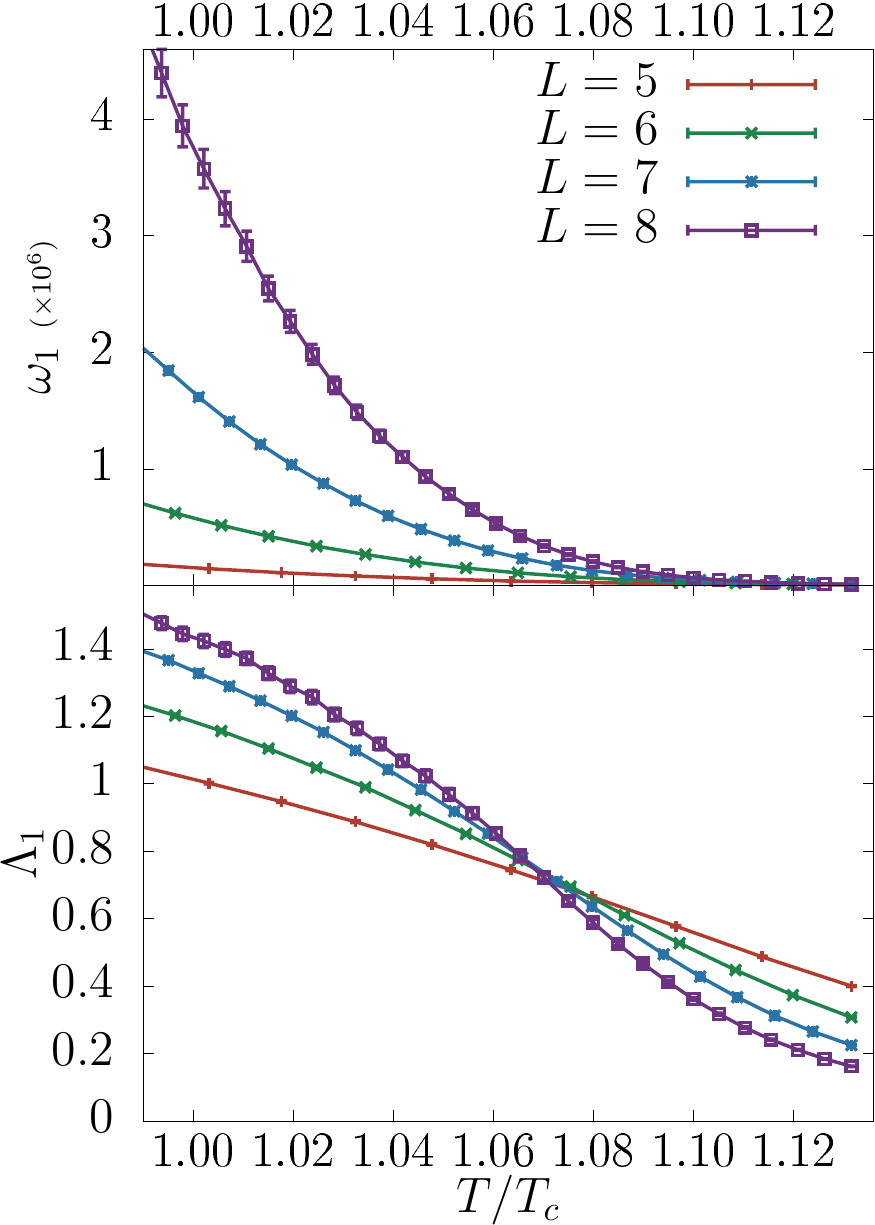}
    \caption{Plot of $\omega_1$ (on the \textbf{top}), Eq. \eqref{eq:omegas}, and the cummulant  $\Lambda_1$ (on the \textbf{bottom}), Eq. \eqref{eq:Lambda_cummulant},  as a function of  $T/T_c$. }
    \label{fig:Lambda1}
\end{figure}

\begin{figure}[h]
    \centering
    \includegraphics[scale=0.75]{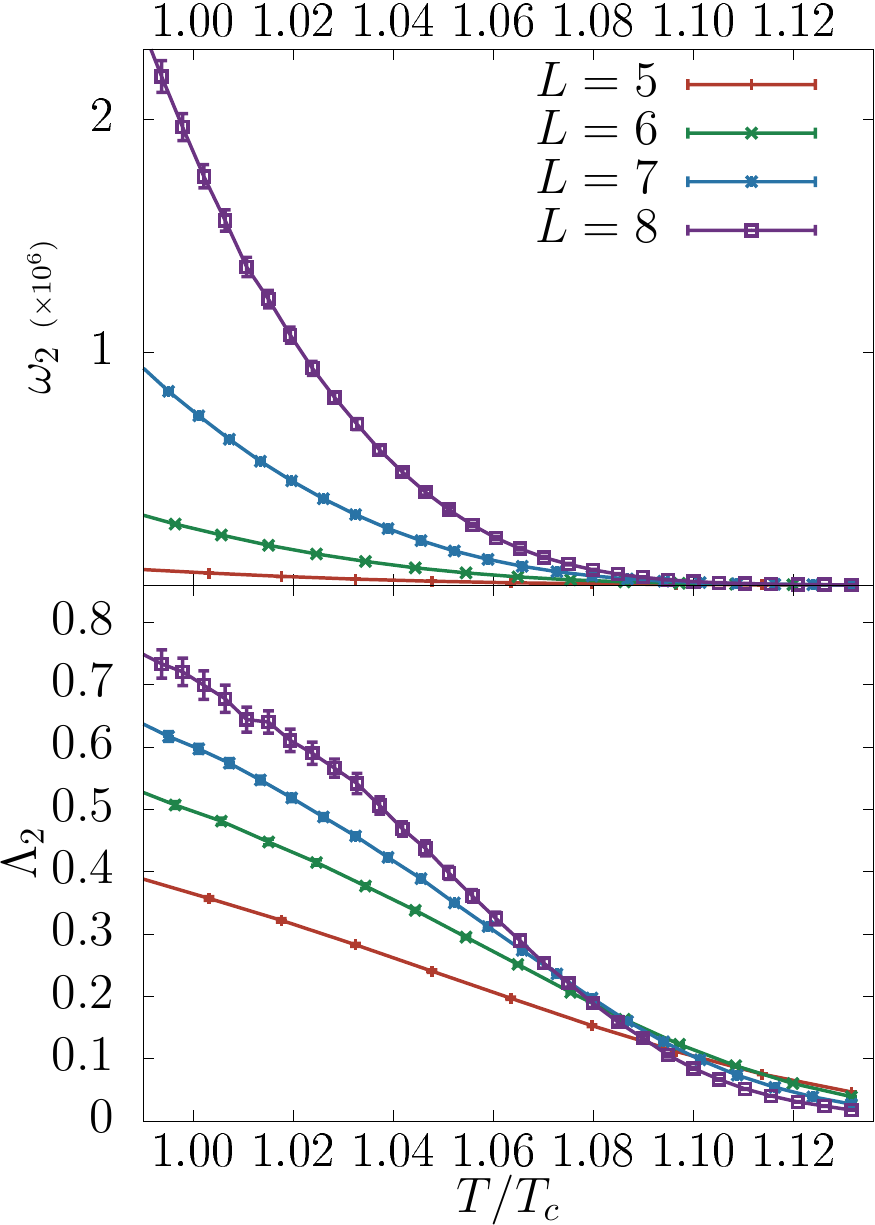}
     \caption{\textbf{Up:} Plot of $\omega_2$ as a function of  $T/T_c$. \textbf{Down:} Plot of $\Lambda_2$ as a function of $T/T_c$.}
    \label{fig:Lambda2}
\end{figure}

In Figs.~\ref{fig:Lambda1} and ~\ref{fig:Lambda2} one can find a two-panel figure for $\omega_1$, $ \Lambda_1$, and for $\omega_2$, $\Lambda_2$, respectively, as a function of $T/T_c$.

Notice that both $\Lambda$-cumulants show crossing points (better signal for $\Lambda_1$), as expected, drifting very slowly to the critical point. 

In addition the connected susceptibilities $\omega_1$ and $\omega_2$ diverge in the critical region as predicted by the theory: 
$\omega_{1,2} \propto |T-T_c|^{-\gamma_3}$ with $\gamma_3=6 \nu- \frac{3}{2} \nu \eta$ in six dimensions~\cite{fernandez:22} ($\gamma_3=3$ for $D=6$ and the MF exponents).

\clearpage

\bibliography{apssamp}

\end{document}